\documentclass[aps,prb,showpacs,twocolumn,latexsym]{revtex4}
\usepackage{amssymb}
\usepackage{latexsym}
\usepackage{graphicx}
\usepackage{amsmath}
\input{tcilatex}
\begin{document}
\title{Hall coefficient in an interacting electron gas}
\date{\today }
\author{ M.~Khodas  and  A.M.~Finkel'stein }
\begin{abstract}
The Hall conductivity in a weak homogeneous magnetic field, $\omega_{c}\tau \ll 1$, 
is calculated. 
We have shown that to leading order in $1/\epsilon_{F}\tau$ the Hall
coefficient $R_{H}$ is not renormalized by the electron-electron
interaction. Our result explains the experimentally observed stability of
the Hall coefficient in a dilute electron gas in Si MOSFETs not too close to
the metal-insulator transition. 
We avoid the currently used procedure that introduces an
artificial spatial modulation of the magnetic field.
The problem of the Hall effect is reformulated in a way such that
the magnetic flux associated with the scattering process becomes the central
element of the calculation.
\end{abstract}
\pacs{72.10.-d, 71.30.+h, 71.10.Ay}
\affiliation{Department of Condensed Matter Physics, the Weizmann Institute of Science,
Rehovot, 76100, Israel}
\maketitle
%
\section{Introduction}
In recent experiments on the two-dimensional dilute electron gases the
renormalization of the Fermi-liquid parameters is found 
to be significant.\cite{Pudalov1}$^{-}$\cite{Prus}
These strong renormalizations are natural for a dilute gas with a large value of
the ratio of the Coulomb energy to the Fermi energy. It was shown, however,
that in a high mobility 
MOSFETs (metal-oxide-semiconductor field-effect transistor) 
the density of electrons $n_{H}$ found from the Hall
coefficient $R_{H}=1/n_{H}ec$ coincides with that obtained from the
Shubnikov$-$de Haas measurements $n$ to within a few percent 
[see Fig.~4 in Ref. \onlinecite{Pudalov2} and the inset in Fig.~4 in Ref. \onlinecite{Shashkin}(b)].
This remarkable fact raises a question about the
electron-electron (\textit{e-e}) interaction renormalizations of the Hall
conductivity $\sigma _{xy}$ in a small magnetic field, $\omega _{c}\tau \ll
1 $, where $\omega _{c}$ is a cyclotron frequency and $\tau $ is the free
path time. Notice that while the \textit{e-e} interaction is very strong, in
other aspects this system is rather simple, because the effect of the
crystalline lattice can be safely studied in the effective-mass
approximation. Apart from impurities, one may consider the electron liquid
as if it is in a translational invariant background. Therefore the
influence of the interaction on the Hall coefficient should not be masked by
unnecessary complications. The issue of the Hall coefficient renormalization
has not been discussed much in the literature despite its obvious
importance. We have in mind the renormalization of the leading in $\tau
_{tr} $ terms in the conductivity tensor, i.e., of the Hall
conductivity $\sigma _{xy}\propto B\tau _{tr}^{2}$, and the diagonal term $%
\sigma _{xx}=ne^{2}\tau _{tr}/m^{\ast }$. Here $\tau _{tr}$ is the transport
relaxation time, and $B$ is a magnetic field; notice that $R_{H}\equiv
B^{-1}\sigma _{xy}/(\sigma _{xx})^{2}$.

In this paper we examine the effect of the \textit{e-e} interaction on the
Hall coefficient within the microscopical theory. We avoid the currently
used procedure that introduces an artificial spatial modulation of the
magnetic field. Based on an idea of the flux of a loop in a diagrammatic
technique, we obtain a gauge invariant procedure for the calculations
in the presence of a homogeneous magnetic field which makes it possible to
conduct the analysis of $\sigma _{xy}$ in a general manner. In particular,
in the course of the calculations the translational invariance of the
problem is maintained, and the usual rather involved procedure of extracting
a constant $B$ from the $\mathbf{ q}\rightarrow 0$ limit of a singular
expression for the vector potential $\mathbf{ A}(\mathbf{ q})$ can be avoided.

In this paper, we show that the Hall coefficient for electrons with the
quadratic spectrum, $\epsilon (p)=p^{2}/2m,$ in the leading order in $\tau
_{tr}$, is not renormalized by the \textit{e-e} interaction, i.e., 
$R_{H}=1/nec$. [In fact, the cancellation of the \textit{e-e} renormalization
in the Hall coefficient holds for any spherically symmetric dispersion $%
\epsilon (|\mathbf{ p}|)$; see Appendix~\ref{Appendix_B} for comments.] This
result explains the stability of the Hall coefficient $R_{H}$ to the 
\textit{e-e} interaction demonstrated in Refs. \onlinecite{Pudalov2} and \onlinecite{Shashkin}(b).
Notice that the Altshuler-Aronov corrections to $\sigma _{xx}$ and $\sigma _{xy}$ 
teach us
that there cannot be any general principle for the absence of
renormalizations in $R_{H}$, because corrections to $R_{H}$ of the order $%
1/\epsilon _{F}\tau _{tr}$ do exist.\cite{Altshuler1} Therefore the fact
stated here about the cancellation of \textit{e-e} renormalizations in the
leading term of $R_{H}$ should be proved specifically.

The renormalization of $\sigma _{xy}$ can be also studied within the
Fermi-liquid theory following the line given in Ref.\cite{PN}. The magnetic
field $\mathbf{ B}$ comes into a transport equation in a combination with the
velocity of a charge carrier as $(e[\mathbf{ v}_{\mathbf{ p}}\times \mathbf{ B}%
]\partial /\partial \mathbf{ p})\delta n_{\mathbf{ p}}$, where $\delta n_{%
\mathbf{ p}}$ is the departure of the distribution function from equilibrium
due to the applied electric field (see e.g., Ref. \onlinecite{Ziman}). In the
presence of the electron-electron interaction one may expect in addition
terms like $(e[\mathbf{ v}_{\mathbf{ p}}\times \mathbf{ B}]\partial /\partial 
\mathbf{ p})\langle f_{\mathbf{ pp}^{\prime }}\delta n_{\mathbf{ p}^{\prime
}}\rangle $, or $\langle f_{\mathbf{ pp}^{\prime }}(e[\mathbf{ v}_{\mathbf{ p}%
^{\prime }}\times \mathbf{ B}]\partial /\partial \mathbf{ p}^{\prime })\langle
f_{\mathbf{ p}^{\prime }\mathbf{ p}^{\prime \prime }}\delta n_{\mathbf{ p}%
^{\prime \prime }}\rangle \rangle $, where $\langle \ldots  \rangle $ means the
average over the Fermi surface. Here $\langle f_{\mathbf{ pp}^{\prime
}}\delta n_{\mathbf{ p}^{\prime }}\rangle $ is the response of the
Fermi liquid to the departure of the distribution function from equilibrium.\cite{PN} 
This approach gives $\sigma _{xy}$ renormalized by the 
\textit{e-e} interaction which when combined with $\sigma _{xx}$ yields the Hall
coefficient $R_{H}$ with no renormalization factors. It is, however, not
clear that the derivation of the Hall conductivity in a weak magnetic field
via the transport equation within the Fermi-liquid theory is accurate
enough. For instance, consider the following process: the electric field
leads to the departure of the distribution function from equilibrium; the
electron liquid responds to this $\delta n$ via the electron interaction;
the magnetic field acts on the result of this response and transforms it
into the Hall current involving again the \textit{e-e} interaction. The 
nonfactorized part of such two time interaction process $\langle \langle (e[%
\mathbf{ v}_{\mathbf{ p}^{\prime }}\times \mathbf{ B}]\partial /\partial 
\mathbf{ p}^{\prime })f_{np}(\mathbf{ p};\mathbf{ p}^{\prime };\mathbf{ p}%
^{\prime \prime })\delta n_{\mathbf{ p}^{\prime \prime }}\rangle \rangle $
corresponds to nonpair correlations, and is beyond the scope of the 
Fermi-liquid theory. (Some doubts about the derivation via the kinetic equation
were expressed in the literature, e.g., Ref. \onlinecite{Ziman}, Sec.~7.12.) 
We will see, however, that this contribution is small by the
parameter $1/\epsilon _{F}\tau _{tr}$. Therefore the present microscopical
derivation justifies the applicability of the Fermi-liquid treatment of this
problem.

In Sec. II the technicalities of the calculation of $\sigma _{xy}$ in a
constant magnetic field are given. In Sec. III the Hall conductivity of the
noninteracting electrons is considered, and in particular a nontrivial
problem of the Hall conductivity in a system with an arbitrary (nonshort
range) impurity potential is studied. By studying this case we make the
needed preparations for the analysis of the effects of the \textit{e-e}
interaction to the Hall conductivity [see Eqs.~(\ref{poly4})$-$(\ref{Fans2})].
The problem of the Hall effect in a weak magnetic field is reformulated in a
way such that the magnetic flux associated with the scattering process
becomes the central element of the calculation. In Sec. IV the
role of the \textit{e-e} interaction is analyzed. There, the main
contribution to $\sigma _{xy}$ is given by Eq.~(\ref{leading}). The latter
expression contains a product of two terms; one describes the scattering of
the quasiparticles by impurities, while the other corresponds to the
decoration of the current vertices by the \textit{e-e} interaction.
The derivatives over the momentum in Eq.~(\ref{leading}) correspond to the vector
product of the two coordinate differences, see Eq.~(\ref{poly4}), in the
representation of the flux which leads to the skew action of the magnetic
field.

In the concluding section for completeness we discuss $R_{H}$ using the
Fermi-liquid theory.\cite{PN} In Appendix~\ref{Appendix_C} we reproduce the
known answer (Ref. \onlinecite{Fukuyama2}) for the weak localization corrections
to $\sigma _{xy}$ as an instructive example of the calculation within the
new procedure.

\section{Diamagnetic and Flux contributions in a homogeneous magnetic
field}

To describe electrons moving in a magnetic field the magnetic vector
potential $\mathbf{ A}$ can be introduced through the extension of the
momentum in the kinetic energy term: 
$\epsilon (\mathbf{ p})\rightarrow \epsilon (\mathbf{ p}-(e/c)\mathbf{ A})$. 
Although the analysis presented in this paper is valid for any spectrum with a spherical
symmetry (see Appendix~\ref{Appendix_B}), 
the consideration in the main text will be limited to the case 
of the electron gas with a quadratic spectrum
\begin{equation}
H_{0}=\frac{1}{2m}\left( \mathbf{ p-}\frac{e}{c}\mathbf{ A}\right) ^{2},
\label{magnetic_vertex}
\end{equation}
where $m$ is the conduction band mass (i.e., $m$ is not
renormalized yet by the electron-electron interaction), and we use the fact
that for low concentrations of the electrons only the quadratic term in $%
\epsilon (p)$ is relevant.

Notice that the homogeneous magnetic field demands special care as the
Fourier components of the vector potential, $\mathbf{ A}(\mathbf{ q})=\int
d^{d}re^{-i\mathbf{ qr}}\mathbf{ A(r)}$, are singular in this important case.
Let us take the vector potential in the Landau gauge, 
\begin{equation}
\mathbf{ B}=B\widehat{\mathbf{ z}};\hspace{10pt}A_{x}=0,\hspace{10pt}A_{y}=Bx.
\label{Landau_gauge}
\end{equation}
Then the Fourier components of $\mathbf{ A(r)}$ are 
\begin{equation}
A_{y}(\mathbf{ q})=iB\frac{\partial }{\partial q_{x}}\delta ^{d}(\mathbf{ q}).
\label{FT_of_Landau_gauge}
\end{equation}
When inserted directly into diagrams the expression for $A_{y}(\mathbf{ q})$
being highly singular leads to complications. In order to circumvent this
difficulty it was proposed in Ref.~\onlinecite{FEW} to introduce a magnetic field
modulated in space at some wave vector $\mathbf{ q}$ in addition to a
homogeneous electric field at finite frequency $\omega ,$ and to study the $(%
\mathbf{ q},\omega )$ response of the current. It has been argued in 
Ref.~\onlinecite{FEW} that in the calculation of $\sigma _{xy}$ one should perform the
limiting procedure as follows: first extract the $\mathbf{ q}$-linear
contributions related to the vector potential and arrange them into the
gauge invariant combination $\mathbf{ B}(\mathbf{ q})=i{\mathbf{ q}\times 
\mathbf{ A}}(\mathbf{ q})$; next let $\mathbf{ q}\rightarrow 0$ keeping $%
\mathbf{ B}(\mathbf{ q})$ to be finite; and, finally, perform the dc limit for
the current response. The first step in this procedure requires combining
different pieces into the gauge invariant combination, and this involves 
nontrivial cancellations amongst different diagrams. The outlined procedure is
effective when a specific process is needed to be calculated 
(see, e.g., Refs.~\onlinecite{Altshuler1} and \onlinecite{Fukuyama2}), but unfortunately it is
unsuitable for a general analysis.

The method proposed below treats the magnetic field at $\mathbf{ q}=0$ 
\textit{ab initio}, and at no stage of the calculation does the vector
potential appear explicitly. In addition to the apparent gauge invariance
the developed technique shows clearly that although the Hamiltonian in the
magnetic field is not translational invariant, any measured quantity like
conductivity can be calculated in an explicitly transitional invariant
manner. The calculation is based on the fact that the Green's function in a
constant magnetic field can be represented in 
the form \cite{Levinson}$^{-}$\cite{Edelstein} 
\begin{equation}
\mathcal{ G }(\mathbf{ r}_{1},\mathbf{ r}_{2},\tau )=\exp \left[ (ie/c)\Phi (%
\mathbf{ r}_{1},\mathbf{ r}_{2})\right] \tilde{{\mathcal G}}(\mathbf{ r}_{1}-
\mathbf{ r}_{2},\tau ),  \label{defGtilda}
\end{equation}
where the phase $\Phi $ is equal to 
\begin{equation}
\Phi (\mathbf{ r}_{1},\mathbf{ r}_{2})=
(\mathbf{ r}_{1}-\mathbf{ r}_{2})
\mathbf{ A }\left( \frac{  \mathbf{ r}_{1}+\mathbf{ r}_{2}}{  2  }\right)  \label{phase1}
\end{equation}
and $\tilde{{\mathcal G}}$ is the core Green's function which contains the
information about the Landau-level quantization. The function $\tilde{%
{\mathcal G}}$ is translational and rotational invariant and is also
invariant with respect to different gauge representations of the magnetic
field $\mathbf{ B}$ (see Appendix \ref{Appendix_A}). Notice that in the
presence of \textit{e-e} interaction the representation~(\ref{defGtilda})
and (\ref{phase1}) is still valid.

The rotational invariance of the core function $\tilde{{\mathcal G}}$ is
responsible for the fact that the Hall current originates either from the
diamagnetic part of the current operator, or with the participation of the
phase terms $\Phi .$ No contribution to $\sigma _{xy}$ comes out from the
core functions alone. This is because the calculation of $\sigma _{xy}$ in
the Kubo formula involves the averaging over the momenta of the nondiagonal
current-current correlator $\langle J_{x}J_{y}\rangle $. Since the functions 
$\tilde{{\mathcal G}}$ are rotational invariant, the angular integration of
the type $\langle p_{x} \ldots p_{y}^{\prime }\rangle $ makes the core function's
contribution to $\sigma _{xy}$  zero if the scattering by impurities is
simple enough and does not itself produce the skew effect on average. Then
only the interference of the phase factor $\Phi $ or the diamagnetic current
make it possible to achieve the skew action of the magnetic field to avoid 
the vanishing of the $\langle J_{x}J_{y}\rangle $ 
current correlator after the angular
integration. (Notice in this connection that the Lorentz force can be
interpreted as a consequence of the phase accumulation in the magnetic
field, Ref.~\onlinecite{Feynman}.)

The diamagnetic part of the current operator, $-(e^{2}/mc)\mathbf{ A,}$ being
nongauge invariant has to be completed into a gauge invariant contribution
in any calculation of a measurable quantity. This completion comes out
naturally here due to the differentiation of the phase factors in the
Green's functions in the current operator, 
\begin{eqnarray}
{\ J}_{\nu }(\mathbf{ r}_{i}) & = & 
\left( e/2mi\right) \lim_{{\mathbf{ r}_{i}^{\prime }}
\rightarrow {\mathbf{ r}_{i}}} (\nabla _{{\mathbf{ r}_{i}^{\prime }}}^{\nu }
-\nabla _{{\mathbf{ r}_{i}}}^{\nu })\psi ^{\dagger }
(\mathbf{ r}_{i})\psi (\mathbf{ r}_{i}^{\prime }) \nonumber \\
 & & -(e^{2}/mc)A^{\nu }(\mathbf{ r}_{i})\psi ^{\dagger }(\mathbf{ r}_{i})\psi (\mathbf{ r}_{i}).
\label{curroper1}
\end{eqnarray}
Consider the current vertex at the point $\mathbf{ r}_{i}$ connected to the
rest of a diagram (e.g., for a current-current correlator) by a pair of
Green's functions ${\mathcal G}(\mathbf{ r}_{1},\mathbf{ r}_{i})$ and ${\mathcal %
G}(\mathbf{ r}_{i},\mathbf{ r}_{2})$. Then the diamagnetic part results in 
\[
(e^{2}/mc)\mathbf{ A}(\mathbf{ r}_{i}){\mathcal G}(\mathbf{ r}_{1},\mathbf{ r}%
_{i}){\mathcal G}(\mathbf{ r}_{i},\mathbf{ r}_{2}),
\]
whereas the first term yields 
\[
(e/2m)[{\mathcal G}(\mathbf{ r}_{1},\mathbf{ r}_{i})(-i)\nabla _{\mathbf{ r}_{i}}%
{\mathcal G}(\mathbf{ r}_{i},\mathbf{ r}_{2})+i\nabla _{\mathbf{ r}_{i}}{\mathcal %
G}(\mathbf{ r}_{1},\mathbf{ r}_{i}){\mathcal G}(\mathbf{ r}_{i},\mathbf{ r}_{2})].
\]
To treat the phase factors in the Green's functions it is useful to split
the $\mathbf{ A}$ term into two half pieces and introduce the following
relations: 
\begin{equation}\label{commute1}
\begin{split}
&[-i\nabla_{\mathbf{ r}_{i}}-(e/c)\mathbf{ A}(\mathbf{ r}_{i})] 
{\mathcal G} (\mathbf{ r}_{i},\mathbf{ r}_{2};\tau ) =    \\
     &\exp \left[ \left( ie/c\right) \Phi (\mathbf{ r}_{i},\mathbf{ r}_{2})\right]  \\
       & \times \left[ -i\nabla_{\mathbf{ r}_{i}}+(e/2c)\left( \mathbf{ r}_{i}-\mathbf{ r}_{2}\right) 
\times \mathbf{ B}\right] \tilde{{\mathcal G}} (\mathbf{ r}_{i}-\mathbf{ r}_{2};\tau ), 
\end{split}
\end{equation}
and 
\begin{equation}\label{commute2}
\begin{split}
&[i\nabla _{\mathbf{ r}_{i}}-(e/c)\mathbf{ A}(\mathbf{ r}_{i})] 
{\mathcal G}(\mathbf{ r}_{1},\mathbf{ r}_{i};\tau ) =\\
&  \exp \left[ \left( ie/c\right) \Phi (\mathbf{r}_{1},\mathbf{r}_{i})\right]  \\
& \times \left[ -i\nabla _{\mathbf{ r}_{1}}-(e/2c)
\left( \mathbf{ r}_{1}-\mathbf{ r}_{i}\right) \times \mathbf{ B}\right] \tilde{{\mathcal G}}(\mathbf{ r}_{1}-%
\mathbf{ r}_{i};\tau ). 
\end{split}
\end{equation}
The terms on the right-hand side that contain $B$ explicitly yield the diamagnetic
contribution extended to gauge invariant combinations. Combining Eqs. (\ref
{commute1}) and (\ref{commute2}) we obtain 
\begin{equation}\label{diamcontr2}
\begin{split}
& J_{extended}^{diam} = (e\omega _{c}/4)\left[ \left( \mathbf{ r}_{i}-\mathbf{ r}%
_{2}\right) \times \widehat{\mathbf{ B}}-\left( \mathbf{ r}_{1}-\mathbf{ r}%
_{i}\right) \times \widehat{\mathbf{ B}}\right] \\
& \times 
\tilde{{\mathcal G}}(\mathbf{ r}_{1}-\mathbf{ r}_{i})\tilde{{\mathcal G}}(\mathbf{ r}_{i}-\mathbf{ r}_{2})
R(\mathbf{ r}_{1}-\mathbf{ r}_{f},\mathbf{ r}_{2}-\mathbf{ r}_{f}),
\end{split}
\end{equation}
where $R(\mathbf{ r}_{1}-\mathbf{ r}_{f},\mathbf{ r}_{2}-\mathbf{ r}_{f})$ is
the remaining part of a correlator ending at a point $\mathbf{ r}_{f}$.
Notice that $J_{extended}^{diam}$ is also translation invariant. In Eq. (\ref
{diamcontr2}) the cyclotron frequency has been introduced in such way that
its sign is that of the carriers: $\omega _{c}=eB/mc$.
The phase factor $\Phi (\mathbf{ r}_{1},\mathbf{ r}_{2})$ appearing in the
Green's function ${\mathcal G}(\mathbf{ r}_{1},\mathbf{ r}_{2},\tau )$ may be
rewritten as the integral over the straight line: 
\begin{equation}
\Phi (\mathbf{ r}_{1},\mathbf{ r}_{2})=(\mathbf{ r}_{2}-\mathbf{ r}_{1})
\mathbf{ A} \left( \frac{ \mathbf{ r}_{1}+\mathbf{ r}_{2} }{ 2 } \right)
     =\int_{\mathbf{ r}_{1}}^{\mathbf{ r}_{2}} \mathbf{ A}(\mathbf{ r})d\mathbf{ r}.  
\label{int_phase}
\end{equation}
Then collecting the phase factors from all Green's functions of a given
loop of a diagram one will obtain the so-called ``flux of a loop"
of a diagram. [The phase factors that appear in the 
right-hand side of
Eqs. (\ref{commute1}) and (\ref{commute2}) after passing through the
differentiation operator join other such factors from the rest of the
Green's functions.] It follows from Eq. (\ref{int_phase}) that the phase
associated with each loop in a diagram is proportional to the flux, 
\begin{equation}
\Phi _{loop}=\oint \mathbf{ A}(\mathbf{ r})d\mathbf{ r},
\end{equation}
with the integral taken along the closed polygon corresponding to the loop.
(The importance of the flux in the calculation of the Hall effect in the
insulating state was emphasized in Refs.~\onlinecite{Holstein} and \onlinecite{Imry}.)
Notice that the polygon is oriented along the loop in the direction of the
arrows of the Green's function. The magnetic flux through the polygon is
gauge and translational invariant. It is equal to $\mathbf{ BS},$ where $%
\mathbf{ S}$ is the oriented area of a polygon. Depending on the situation
one can decompose the polygon into the sum of different pieces. The
decomposition can be done in a variety of ways, and a convenient choice may
simplify the calculation. This is an advantage of this method. With the
purpose of a general analysis of the Hall conductivity we will decompose the
loop polygon in a diagram for the current-current correlator into the sum of
triangles, and then use the fact that the flux through a triangle $(\mathbf{ r%
}_{1},\mathbf{ r}_{2},\mathbf{ r}_{3})$ is equal to 
$(1/2)\mathbf{ B}\cdot [  ( \mathbf{ r}_{2}-\mathbf{ r}_{1})\times (\mathbf{ r}_{3}-\mathbf{ r}_{2}) ]$. 
The details of the decomposition procedure will be postponed to the next section
[see Eqs. (\ref{poly1})$-$(\ref{poly4})], where it will be shown how the flux
contribution can be analyzed for the calculation of the Hall conductivity in
a system with an arbitrary (nonshort range) impurity potential.

Let us summarize. It follows from the symmetry of the problem that
the Hall current may appear either from the diamagnetic part of the current
operator, or through the phase factors of the Green's functions. Partially
we have used the phase factors to extend the diamagnetic term in Eqs. 
(\ref{commute1})$-$(\ref{diamcontr2}). The rest of the phase factors contributions
can be organized in the form of the fluxes of the loops. The main
consequence of this structure is that if one is interested in $\sigma _{xy}$
linear in the external magnetic field only, $\sigma _{xy}\propto B$, then it
is enough for the calculations of $\sigma _{xy}$ that the core Green's
functions will be taken in the limit $B\rightarrow 0$, when $\tilde{{\mathcal %
G}}$ coincide with the Green's functions in the absence of the magnetic
field: $\tilde{{\mathcal G}}_{B\rightarrow 0}={\mathcal G}_{B=0}$. The latter
fact enables us to reduce the calculation of $\sigma _{xy}$ in a small
magnetic field to a diagrammatic problem that uses the Green's functions in
the absence of the magnetic field.
\section{Hall conductivity for noninteracting electrons in the presence
of a finite range disorder}

We first illustrate the use of Eq. (\ref{diamcontr2}) by deriving the Drude
Hall conductivity for the simplest case of short-range impurities. The
conductivity will be calculated in the framework of the Kubo 
linear-response
theory. The current-current correlator on the Matsubara frequencies is
defined as

\begin{equation}
\Pi _{\alpha \beta }(i\omega _{n})=-\int_{0}^{\beta }e^{i\omega _{n}\tau}
\langle  T_{\tau }J_{\alpha }(\tau )J_{\beta }(0) \rangle.  
\label{correlator}
\end{equation}
The analytic continuation from the discrete frequencies on the upper complex
half plane to the real axes yields the retarded correlator $\Pi _{\alpha
\beta }^{R}(\omega )$. This correlator is directly related to the
conductivity, 
\begin{equation}
\sigma_{\alpha \beta } (\omega )=\frac{i}{\omega }\Pi _{\alpha \beta }^{R}(q=0,\omega ).
\label{conductivity1}
\end{equation}

In the case of a short-range disorder the correlator (\ref{correlator}) in
the Drude approximation is equal to 
\begin{widetext}
\begin{equation}\label{Drude1}
\begin{split}
\Pi _{xy}(r_{i},r_{f};\tau ) & =  
(e^{2}/4m^{2}) 
\bigglb( 
{ \lim_{{\mathbf{ r}_{i}^{\prime }}\rightarrow {\mathbf{ r}_{i}}}}
\{
 -i{\nabla _{{\mathbf{ r}_{i}^{\prime }}}^{x}}-(e/c)A^{x} ({\mathbf{ r}_{i}^{\prime }}) 
\}
+ i{\nabla }_{{\mathbf{ r}_{i}}}^{x}-(e/c)A^{x}({\mathbf{ r}_{i}}) \biggrb)
\bigglb(  -i\nabla _{{\mathbf{ r}_{f}}}^{y}-(e/c)A^{y}({\mathbf{ r}_{f}}) 
\\
   &   +    
\lim_{{\mathbf{ r}_{f}^{\prime }}\rightarrow {\mathbf{ r}_{f}}}  
\{ 
i\nabla _{{\mathbf{ r}_{f}^{\prime }}}-(e/c)A^{y}({\mathbf{ r}_{f}^{\prime }}) 
\}
\biggrb) 
\tilde{{\mathcal G}}({\mathbf{ r}_{f}-\mathbf{ r}_{i}},-\tau )
\exp \left[  (ie/c)\Phi ({\mathbf{ r}_{f},\mathbf{ r}_{i}})  \right]  
 \tilde{{\mathcal G}}
({\mathbf{ r}_{i}^{\prime }-\mathbf{ r}_{f}^{\prime }},\tau )
\exp  \left[  (ie/c)\Phi ({\mathbf{ r}_{i}^{\prime },\mathbf{ r}_{f}^{\prime }})   \right] .
\end{split}
\end{equation}
Using Eqs. (\ref{commute1}) and (\ref{commute2}), Eq. (\ref{Drude1}) can be
rewritten as 
\begin{equation}\label{Drude2}
\begin{split}
\Pi _{xy}(\mathbf{ r}_{i},\mathbf{ r}_{f};\tau ) &  = 
(e^{2}/4m^{2})
\bigglb( 
{\lim_{{\mathbf{ r}_{i}^{\prime }}\rightarrow {\mathbf{ r}_{i}}}}
\{ -i{\nabla _{{\mathbf{ r}_{i}^{\prime }}}^{x}}
+
(e/2c)[({\mathbf{ r}_{i}^{\prime }-\mathbf{ r}_{f}^{\prime }})\times \mathbf{ B}]_{x} 
\}
+  i{\nabla }_{{\mathbf{ r}_{i}}}^{x}-(e/2c)[(\mathbf{ r}_{f}-\mathbf{ r}_{i})\times \mathbf{ B}]_{x} 
\biggrb)   
\bigglb(  
-i{\nabla }_{{\mathbf{ r}_{f}}}^{y} 
\\
      &    + 
(e/2c)[(\mathbf{ r}_{f}-\mathbf{ r}_{i})\times \mathbf{ B}]_{y} 
+  {\lim_{{\mathbf{ r}_{f}^{\prime}}\rightarrow {\mathbf{ r}_{f}}}} 
\{
 i{\nabla }_{{\mathbf{ r}_{f}^{\prime }}}^{y}-(e/2c)[({\mathbf{ r}_{i}^{\prime }}-{\mathbf{ r}_{f}^{\prime }})\times 
\mathbf{ B}]_{y} 
\}
\biggrb)   
\tilde{{\mathcal G}}(\mathbf{ r}_{f}-\mathbf{ r}_{i},-\tau )
\tilde{{\mathcal G}}({\mathbf{ r}_{i}^{\prime }}-{\mathbf{ r}_{f}^{\prime }},\tau ).
\end{split}
\end{equation}
\end{widetext}
Notice that when in Eq. (\ref{Drude2}) the gradients from the $J_{x}$ vertex
act on the coordinate dependent terms in the $J_{y}$ part the corresponding
contributions cancel out. The part linear in $B$ arising from the $J_{x}$ 
vertex is equal to 
\begin{equation}\label{Drude2x}
\begin{split}
& \Pi _{xy}^{(x)}(\mathbf{ r}_{i},\mathbf{ r}_{f};\tau ) =  (e^2 \omega_c/ 4 m)  
[(\mathbf{ r}_{i}-\mathbf{ r}_{f})\times \mathbf{ \hat{B} }]_{x}
\\
 &  \quad  \times  
\{ 
[ -i\nabla _{{\mathbf{ r}_{f}}}^{y}
\tilde{{\mathcal G}}(\mathbf{ r}_{f}-\mathbf{ r}_{i},-\tau )] 
\tilde{{\mathcal G}}(\mathbf{ r}_{i}-\mathbf{ r}_{f},\tau )
\\
 &  \quad    +
\tilde{{\mathcal G}}(\mathbf{ r}_{f}-\mathbf{ r}_{i},-\tau ) 
[ i\nabla _{{\mathbf{ r}_{f}}}^{y}
\tilde{{\mathcal G}}(\mathbf{ r}_{i}-\mathbf{ r}_{f},\tau ) ] 
\} ,
\end{split}
\end{equation}
and the other term from the $J_{y}$ vertex is 
\begin{equation}\label{Drude2y}
\begin{split}
 & \Pi _{xy}^{(y)}(\mathbf{ r}_{i},\mathbf{ r}_{f};\tau )  =  (e^2 \omega_c/ 4 m)
[(\mathbf{ r}_{f}-\mathbf{ r}_{i})\times \mathbf{ \hat{ B} }]_{y}
\\
 &  \quad  \times  
\{ 
[ i\nabla _{{\mathbf{ r}_{i}}}^{x}
\tilde{{\mathcal G}}(\mathbf{ r}_{f}-\mathbf{ r}_{i},-\tau ) ] 
\tilde{{\mathcal G}}(\mathbf{ r}_{i}-\mathbf{ r}_{f},\tau )
\\
 &  \quad    +
\tilde{{\mathcal G}}(\mathbf{ r}_{f}-\mathbf{ r}_{i},-\tau ) 
[ -i\nabla _{\mathbf{ r}_{i}}^{x}\tilde{{\mathcal G}}(\mathbf{r}_{i}-\mathbf{ r}_{f},\tau )] 
\}.   
\end{split}
\end{equation}
[Notice that Eqs. (\ref{Drude2x}) and (\ref{Drude2y}) reproduce the structure of
Eq. (\ref{diamcontr2}).] Now the transition $\tilde{{\mathcal G}}%
_{B\rightarrow 0}={\mathcal G}_{B=0}$ can be performed in these expressions,
and as all terms are translational invariant one may use the Fourier
transformation. Then the coordinate difference $(\mathbf{ r}_{f}-\mathbf{ r}%
_{i})$ leads to the differentiation of the Green's function with respect to
momentum: 
\begin{equation} \label{DrudexF}
\begin{split}
 & \Pi _{xy}^{(x)}(q =0,i\omega _{n})=i(e^{2}\omega _{c}/4m)T\sum_{i\epsilon _{n}}\int \frac{d^{d}p}{(2\pi )^{d}}
\\
\times &
[ {\mathcal G}^{A}(\mathbf{ p})\partial /\partial p_{y}
{\mathcal G}^{R}(\mathbf{ p})
- {\mathcal G}^{R}(\mathbf{ p})\partial /\partial p_{y}{\mathcal G}^{A}(\mathbf{ p}) ] {p}_{y}, 
\end{split}
\end{equation}
and correspondingly 
\begin{equation} \label{DrudeyF}
\begin{split}
 & \Pi _{xy}^{(y)}(q = 0,i\omega _{n}) = i(e^{2}\omega _{c}/4m)T\sum_{i\epsilon _{n}}\int \frac{d^{d}p}{(2\pi )^{d}}
\\
 \times & p_{x}[ {\mathcal G}^{A}(\mathbf{ p})\partial /\partial p_{x}{\mathcal G}^{R}(\mathbf{ p})
-{\mathcal G}^{R}(\mathbf{ p})\partial /\partial p_{x}{\mathcal G}^{A}(\mathbf{ p}) ] .  
\end{split}
\end{equation}
The summation over the frequencies and momentum in Eqs. (\ref{DrudexF}) and 
(\ref{DrudeyF}) leads to the standard answer: $\sigma _{xy}=\omega _{c}\tau
\sigma _{xx}$. This part of the calculation is close to the one presented in
Ref.~\onlinecite{Edelstein}. 

Next we consider a nontrivial problem of the calculation of the Drude Hall 
conductivity for noninteracting electrons in the presence of a finite range disorder, 
i.e., for nonshort-range impurities. To the leading order in 
$1/\epsilon_{F} \tau$ the conductivity in this case is given   
by a set of ladder diagrams.
To get $\sigma _{xy}$ that is linear in the external magnetic
field we consider separately the diamagnetic (extended) contribution, and
the flux contribution that comes out from the phase factors accumulated by
the Green's functions in a diagram. In the case of a finite range disorder
the fermion loop does not degenerate to two retraced paths and the flux term
becomes absolutely essential. (Naturally, we have to expand 
$\exp \left[ \left( ie/c\right) \Phi _{loop}\right] $ and keep the linear term only as we
are interested in $\sigma _{xy}$ that is linear in $B$.)

\begin{figure}[h]
\centerline{
    \includegraphics[width=0.45\textwidth]{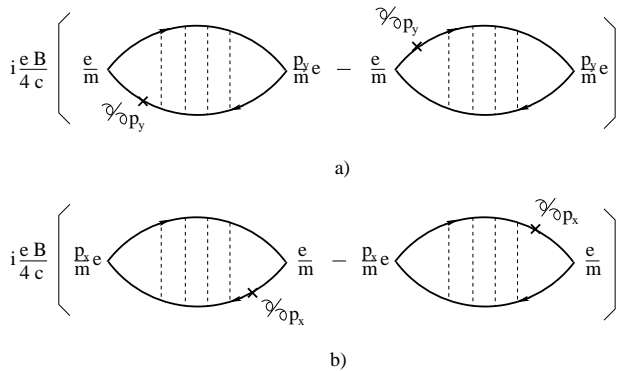}}
\caption{ The diamagnetic contributions: (a) from the $J_{x}$ vertex, (b) from
the $J_{y}$ vertex. }
\label{diam}
\end{figure}

The diamagnetic contribution is represented in Fig.~\ref{diam}. The cross
means differentiation with respect to momentum. Only the Green's functions
adjacent to the current vertex are differentiated. This contribution is
obtained using the relations (\ref{commute1}), (\ref{commute2}), and 
(\ref{diamcontr2}) similar to the case of the short-range impurities. 

\begin{figure}[h]
\centerline{
    \includegraphics[width=0.3\textwidth]{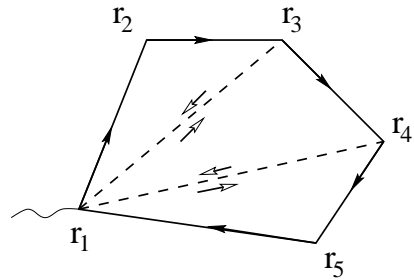}}
\caption{ Flux contribution. }
\label{polygon}
\end{figure}

To analyze the flux contribution to the Hall conductivity we decompose the
loop polygon in a diagram for the current-current correlator into the sum of
triangles. The decomposition will be done as follows: one of the vertexes of
a $n$-vertex polygon is chosen as a reference point (let us call it $\mathbf{ %
r}_{1})$, and then $\mathbf{ r}_{1}$ is connected to all other vertexes. In
result the polygon is decomposed into oriented triangles all having the
reference point as their vertex. (As an example consider a loop
presented in Fig.~\ref{polygon}.) 
The flux through the polygon is therefore equal to the sum of the fluxes
through each triangle: 
\begin{equation}\label{poly1}
\begin{split}
\Phi _{loop} & =\frac{1}{2}\mathbf{ B}\cdot [ (\mathbf{ r}_{2}- \mathbf{ r}_{1})
\times (\mathbf{ r}_{3}-\mathbf{ r}_{2})
\\
 & +(\mathbf{ r}_{3}-\mathbf{ r}_{1})\times (\mathbf{ r}_{4}-\mathbf{ r}_{3})+\cdots ] , 
\end{split}
\end{equation}
which in turn is equal to 
\begin{equation} \label{poly2}
\begin{split}
 \Phi _{loop}  = & \frac{1}{2}\mathbf{ B}\cdot [ (\mathbf{ r}_{2}-\mathbf{ r}_{1})
\times (\mathbf{ r}_{3}-\mathbf{ r}_{2})
 +   (\mathbf{ r}_{3}-\mathbf{ r}_{2})\times (\mathbf{ r}_{4}-\mathbf{ r}_{3})
\\
 + & (\mathbf{ r}_{2}-\mathbf{ r}_{1})
\times (\mathbf{ r}_{4}-\mathbf{ r}_{3})+\cdots ] . 
\end{split}
\end{equation}
Here we rewrote the vector differences $(\mathbf{ r}_{j}-\mathbf{ r}_{1})$ as 
$[(\mathbf{ r}_{j}-\mathbf{ r}_{j-1})+(\mathbf{ r}_{j-1}-\cdots )+(\mathbf{ r}%
_{2}-\mathbf{ r}_{1})]$. 
Finally we add to $\Phi _{loop}$ a zero flux expression that does not change its value, 
\begin{equation}\label{poly3}
\begin{split}
0=\frac{1}{2}\mathbf{ B}\cdot ( \mathbf{ r}_{n}-\mathbf{ r}_{n-1} 
& + \mathbf{ r}_{n-1}-\cdots -\mathbf{ r}_{2}
\\
& +\mathbf{ r}_{2}-\mathbf{ r}_{1} ) \times
\left( \mathbf{ r}_{1}-\mathbf{ r}_{n}\right ) .
\end{split}
\end{equation}
Now we are ready to state the general rule. After some reference point has
been chosen the flux of the oriented polygon is given by 
\begin{equation}
\Phi _{loop}=\frac{1}{2}{\sum_{i\neq j}}^{\prime }(\pm )\mathbf{ B}\cdot \left( 
\mathbf{ r}_{i+1}-\mathbf{ r}_{i}\right) \times \left( \mathbf{ r}_{j+1}-%
\mathbf{ r}_{j}\right) ,
\label{poly4}
\end{equation}
where prime means that each pair $(i,j),$ or $(j,i)$ which is the same,
enters the sum only once, and the pair is oriented with respect to the
reference point in such a manner that its sign is ``$+$'' if we do not pass
the reference point moving along the loop from $r_{i}$ to $r_{j}$ in the
arrow direction. The sign is ``$-$'' otherwise.

After the flux contribution has been decomposed in pairs each pair supplies
the diagram with a vector product of the two coordinate differences. 
When Fourier transformed these coordinate factors lead to the differentiation of
the corresponding Green's function with respect to the momentum components $%
p_{x}$ and $p_{y}$. There are also situations when one or two of $(\mathbf{ r}%
_{i+1}-\mathbf{ r}_{i})$ are adjacent to the current vertexes. Then the
Fourier transformation leads, in addition to the differentiation of the
vertex-attached Green's functions, also to the differentiation of the vertex
momentum.

\begin{figure}[h]
\centerline{
    \includegraphics[width=0.45\textwidth]{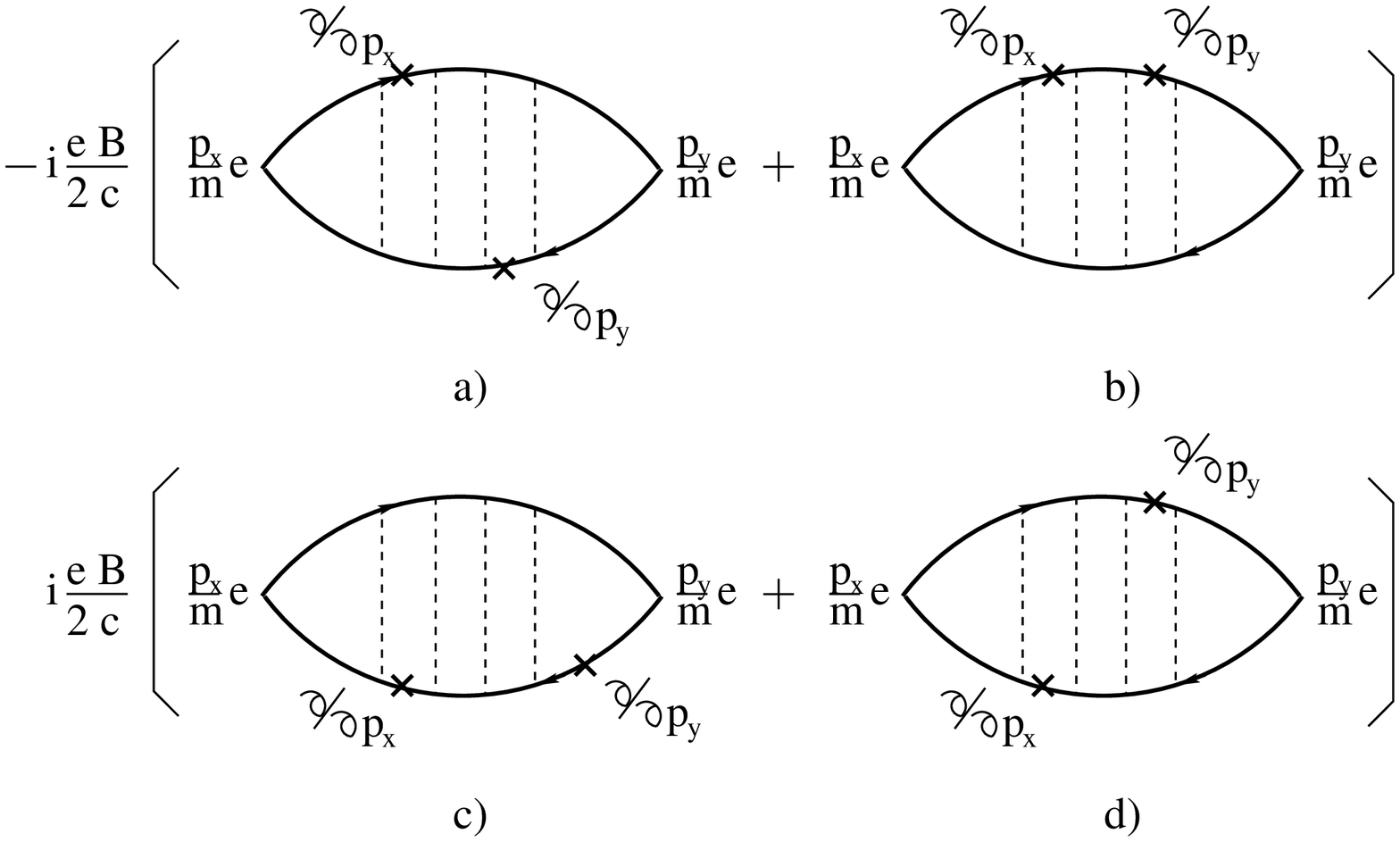}}
\caption{ Flux contributions with no vertex differentiation involved and $%
\partial /\partial p_{x}$ is closer to $p_{x}$ vertex than 
$\partial/\partial p_{y}$.}
\label{flux_dx_dy}
\end{figure}

\begin{figure}[h]
\centerline{
    \includegraphics[width=0.45\textwidth]{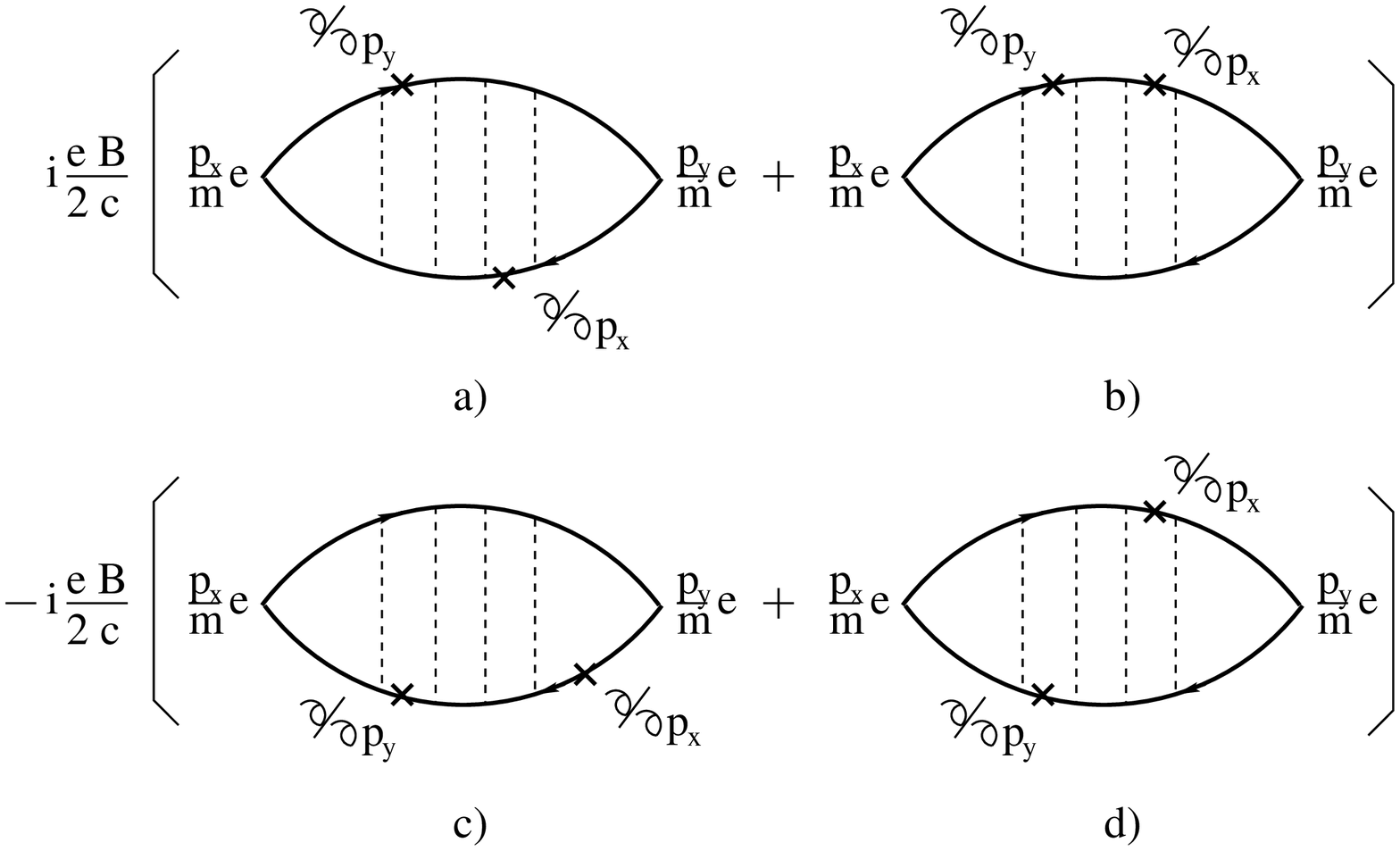}}
\caption{ Flux contributions with no vertex differentiation involved and 
$\partial /\partial p_{y}$ is closer to $p_{x}$ vertex than 
$\partial/\partial p_{x}$.}
\label{flux_dy_dx}
\end{figure}

Let the left-end current vertex be the $p_{x}$ vertex, and let the 
$p_{y}$ vertex be on the right end. The left-end located vertex will be chosen as
the reference point with respect to which the orientation in Eq. (\ref{poly4}%
) has been performed. First consider the terms represented in 
Figs.~\ref{flux_dx_dy} and \ref{flux_dy_dx} when the vertices have not been
differentiated. 
\begin{figure}[h]
\centerline{
    \includegraphics[width=0.45\textwidth]{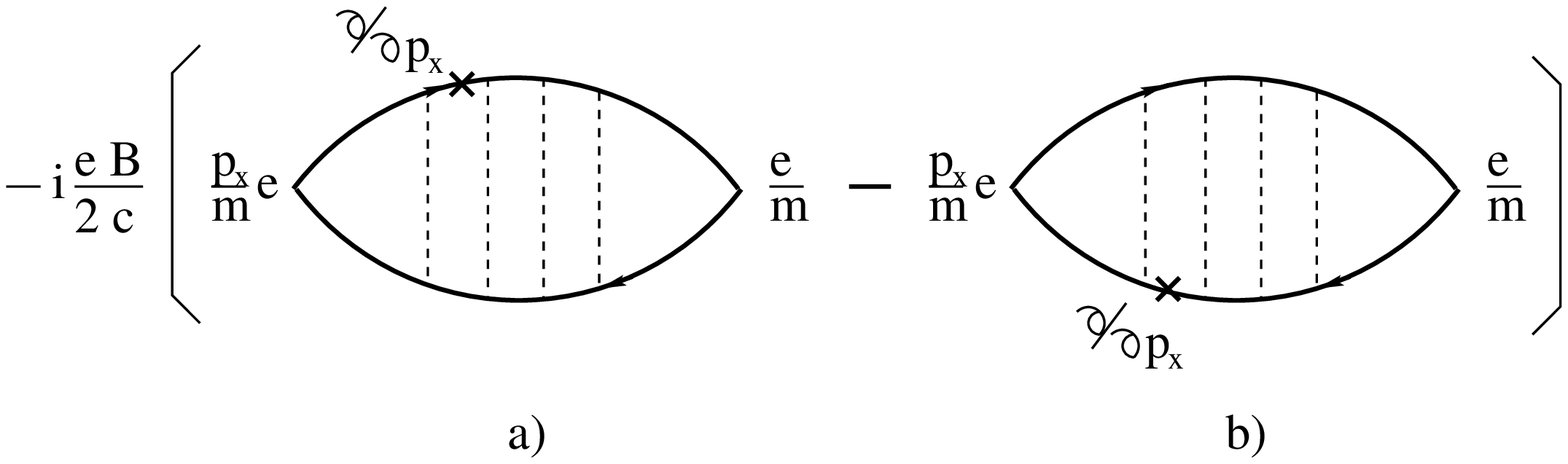}}
\caption{ Flux contributions with $p_{y}$ vertex differentiation involved. }
\label{flux_dpy}
\end{figure}
\begin{figure}[h]
\centerline{
    \includegraphics[width=0.45\textwidth]{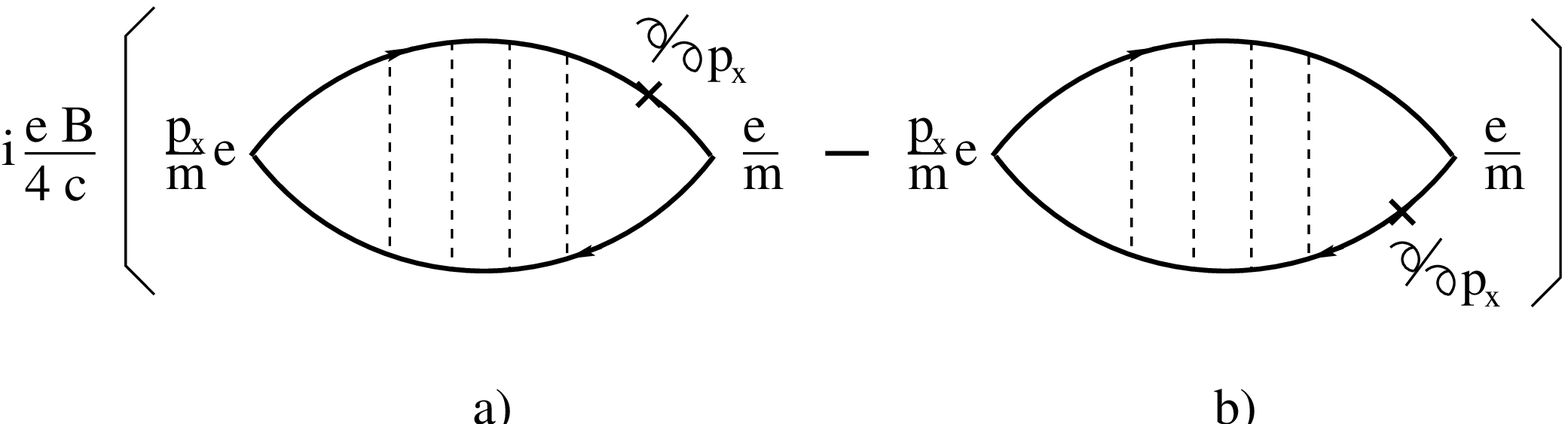}}
\caption{ Compensation of the false contributions of Fig.~\ref{flux_dpy}. }
\label{compensation}
\end{figure}
\begin{figure}[h]
\centerline{
    \includegraphics[width=0.45\textwidth]{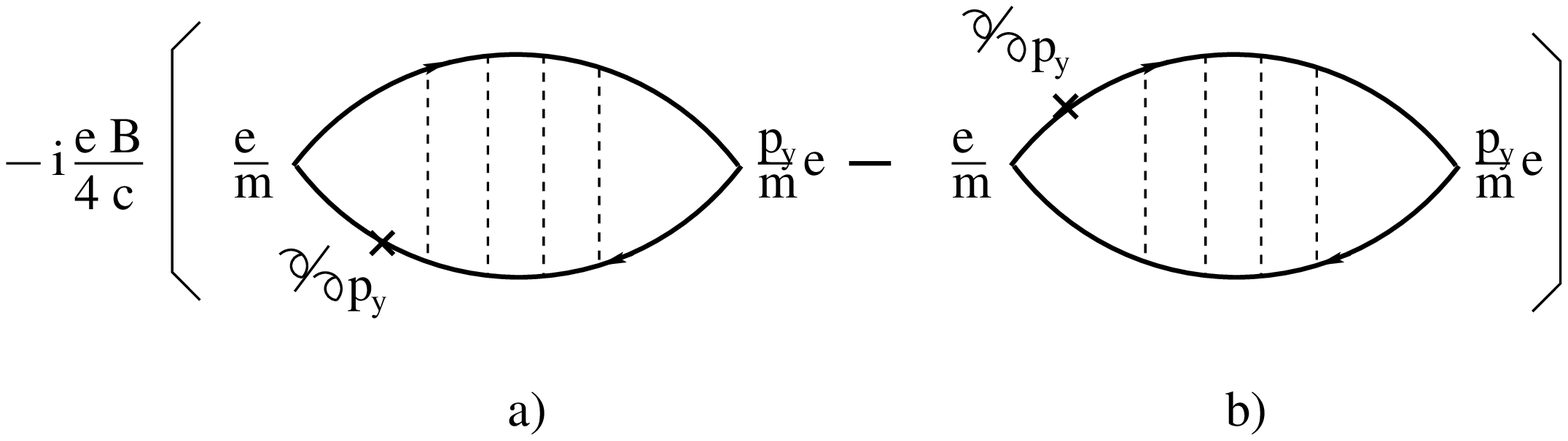}}
\caption{ Flux contributions with $p_{x}$ vertex differentiation involved. }
\label{flux_dpx}
\end{figure}
The difference between the two terms is that in Fig.~\ref{flux_dx_dy} the
differentiation $\partial /\partial p_{x}$ is closer to the left 
$p_{x}$ vertex, whereas in Fig.~\ref{flux_dy_dx} it is $\partial /\partial p_{y}$
that is closer to the $p_{x}$ vertex. The ambiguous terms with $\partial
/\partial p_{x}$ and $\partial /\partial p_{y}$ standing on equal distance, 
i.e., when they stand opposite to each other, are canceled out.
Figure~\ref{flux_dpy} represents schematically the flux contribution involving
the differentiation of the $p_{y}$ vertex at the right-end. 
In Figs.~\ref{flux_dpy}(a) and (b) the $\partial /\partial p_{x}$
differentiation acts on each place in the loop. It should be noticed,
however, that the momentum of the vertex originates from the gradients of
the two Green's functions that are attached to the vertex 
[see, e.g., Eq.~(\ref{Drude1})]. 
Therefore the differentiations in Fig.~\ref{flux_dpy} contain
false nonflux contributions when $\partial /\partial p_{x}$ and $\partial
/\partial p_{y}$ act on the same Green's function attached to the vertex.
The compensation of the false terms is given in Fig.~\ref{compensation}
where only the Green's functions closest to the $p_{y}$ vertex are
differentiated. Finally, in Fig.~\ref{flux_dpx} the contribution involving
the $p_{x}$ vertex differentiation is represented. Here after cancellations
amongst various terms combined into a full derivative only the two terms
with the derivatives that act only on the Green's function closest to the $%
p_{x}$ vertex survive. The apparent asymmetry between Figs.~\ref{flux_dpy}
and~\ref{compensation} compared to Fig.~\ref{flux_dpx} is due to the fact
that it was the $p_{x}$ vertex that was chosen here as a reference point for
the orientation of the pairs, rather than the $p_{y}$ vertex.

\textit{The total contribution. }Let us collect all the terms together.
Notice the cancellations of Fig.~\ref{compensation} with 
Fig.~\ref{diam}(b), and Fig.~\ref{flux_dpx} with Fig.~\ref{diam}(a). 
Then Figs.~\ref{flux_dx_dy}(a), \ref{flux_dx_dy}(b), and 
\ref{flux_dpy}(a) on one hand and 
Fig.~\ref{flux_dx_dy}(c), \ref{flux_dx_dy}(d), and \ref{flux_dpy}(b) on the other
can be packed in combinations such that everything on the right from the 
$p_{x}$ derivative becomes a $p_{y}$ derivative of the current operator: 
\begin{equation}\label{Fans1}
\begin{split}
\Pi _{xy}^{(3,5)}= & i\left( \omega _{c}me^{2}/2\right) [{\mathcal G}^{A}(%
\mathbf{ p}) \partial /\partial p_{x} {\mathcal G}^{R}(\mathbf{ p})
\\
& -
{\mathcal G}^{R} (\mathbf{ p})\partial/\partial p_{x} {\mathcal G}^{A}(\mathbf{ p})]
J_{x} \partial /\partial p_{y} J_{y}.
\end{split}
\end{equation}
(In the last expression we have omitted for brevity 
$T\sum_{i\epsilon_{n}}\int [d^{d}p / (2\pi )^{d}]$, 
as well as in Eqs.~(\ref{Fans2}) and (\ref{Fans3}) below.) 
In addition, Figs.~\ref{flux_dy_dx}(a) and (b) on one hand 
and Fig.~\ref{flux_dy_dx}(c) and (d) on the other yield a similar combination, 
\begin{equation}\label{Fans2}
\begin{split}
\Pi _{xy}^{(4)}= & -i\left( \omega _{c}me^{2}/2\right) 
[{\mathcal G}^{A}(\mathbf{ p}) \partial /\partial p_{y} {\mathcal G}^{R}(\mathbf{ p})
\\
& -
{\mathcal G}^{R}(\mathbf{ p})  \partial /\partial p_{y} {\mathcal G}^{A}(\mathbf{ p})]
J_{x} \partial/\partial p_{x} J_{y}.
\end{split}
\end{equation}
Notice that finally we get a very simple rule: start with the derivative
that stands at the left, and organize all the rest on the right as a
derivative of the $J_{y}$ current; arrange the signs depending on whether
the derivative is on the left-to-right or the right-to-left segments of the
loop, and depending on whether it is a $\partial /\partial p_{x}$ or $%
\partial /\partial p_{y}$ derivative [the change of the sign is in accord
with the orientation rules for pairs formulated for Eq.~(\ref{poly4})].
Remarkably, the diamagnetic and flux terms match each other to produce a
rather simple form. (The Peierls substitution for the hopping matrix
elements in the presence of a magnetic field is useful to discuss the
diamagnetic contribution on an equal footing with the flux term.\cite{Holstein}
In the continuous limit both terms originate from the phase
factor $-ie/c\int \dot{\mathbf{ r}}\mathbf{ A}$ in the path-integral
formulation of the motion of the electron in the presence of the vector
potential.\cite{Feynman_Hibbs}) 

The Onsager relation, $\sigma_{xy}( \mathbf{ B} ) = \sigma_{yx}(- \mathbf{ B} )$,
follows directly from the rule formulated above.
To get $\sigma_{yx}(- \mathbf{ B} )$ from a given contribution to 
$\sigma_{xy}( \mathbf{ B} )$ one has to interchange derivatives 
$\partial  / \partial p_x$ and $\partial  / \partial p_y $
and reverse $ \mathbf{ B }$ to compensate the change of the sign.

Now we use a usual trick, \cite{FEW} and interchange in the $%
\Pi _{xy}^{(4)}$ integration variables $p_{x}$ and $p_{y}$. This leads to a
convenient expression for $\Pi _{xy}$: 
\begin{equation} \label{Fans3}
\begin{split}
\Pi _{xy}= & i\omega _{c}m\frac{e^{2}}{2}\left( {\mathcal G}^{A}(\mathbf{ p})\frac{\partial }{\partial p_{x}}
{\mathcal G}^{R}(\mathbf{ p})-{\mathcal G}^{R}(\mathbf{ p})\frac{\partial }{\partial p_{x}}
{\mathcal G}^{A}(\mathbf{ p})\right )
\\
& \times
\left( J_{x}\frac{\partial }{\partial p_{y}}J_{y}-J_{y}\frac{\partial }{%
\partial p_{y}}J_{x}\right) .
\end{split}
\end{equation}
The scattering by long-ranged impurities results in the renormalization of
the current vertex by the ratio of the transport time $\tau _{tr}$ to the
single-particle scattering time $\tau $: 
\begin{equation}
J_{\alpha }=\frac{\tau _{tr}(|\mathbf{ p}|)}{\tau }\frac{p_{\alpha }}{m}.
\end{equation}
Fortunately, in the combination $(J_{x}\partial /\partial
p_{y}J_{y}-J_{y}\partial /\partial p_{y}J_{x})$ the terms of the type $%
\partial /\partial p_{y}\tau _{tr}(|\mathbf{ p}|)$ cancel each other out.
Then the standard integrations in $\Pi _{xy}$ lead us to a natural
conclusion that the answer for the finite-range impurities differs from that
one for the short-range case by the substitution $\tau \rightarrow \tau
_{tr} $: 
\begin{equation}
\sigma _{xy}=\omega _{c}\tau _{tr}\sigma _{xx}.  \label{relation}
\end{equation}
Correspondingly, it follows from this conclusion that the Hall coefficient
is independent on the impurity range. [Here one point remains to be
cleared up. In the presence of disorder the Green's functions acquire a
self-energy part whose imaginary part is $1/\tau $. Its real part shifts the
chemical potential and hence is of no interest, but its dependence on the
momentum near the Fermi energy may influence the result of the integration
over momentum in Eq.~(\ref{Fans3}). However, in the absence of a special
structure of the scatterers, like in liquid metals, the sensitivity to the
energy of the electron state is very small when $\epsilon _{F}\tau _{tr}\gg
1,$ and this effect can be ignored.]

\begin{figure}[h]
\centerline{\includegraphics[width=0.45\textwidth]{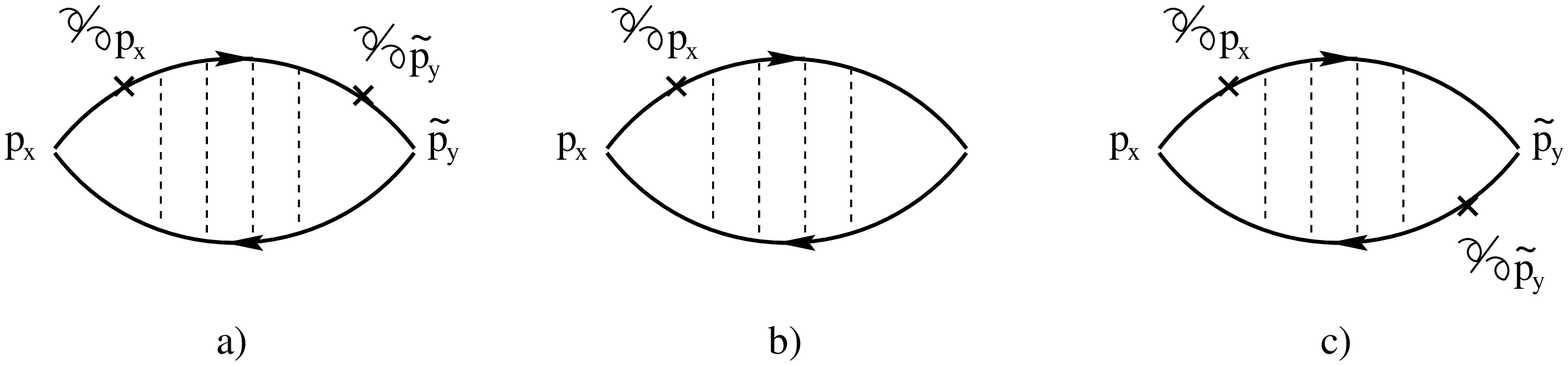}} 
\caption{Cancellation of the singular terms.}
\label{diffusion}
\end{figure}

A comment may be in place here. It was important in the above derivation to
keep the $p$ derivatives in the form of combined expressions of the type $%
\partial /\partial p_{y}J_{y}$ rather than to consider separate terms.
Separately these terms are larger or even more singular than their sum. Let
us see how it works in a somewhat pathological but instructive example of
short-range impurities. In this case the dressing of the vector vertex by
the impurity lines does not give any effect because of the momentum
averaging. The $p_{y}$ derivative makes the vector averaging noneffective
as it is illustrated in Fig.~\ref{diffusion}. 
Each of the three diagrams presented in Fig.~\ref{diffusion} contains the
singular propagator of the diffusion ladder diagrams (diffuson) that does
not vanish after the averaging. This does not lead to any complications,
however, as the sum of these singular terms vanishes identically. Indeed,
the expression corresponding to the ending block (i.e., separated
by the last vertical impurity line) when summed over the three diagrams
turns out to be a derivative $\partial /\partial \widetilde{p}_{y}\left( 
{\mathcal G}^{R}\widetilde{p}_{y}{\mathcal G}^{A}\right) ,$ where $\widetilde{%
\mathbf{ p}}$ is the momentum circulating inside this block. The integration
over $\widetilde{\mathbf{ p}}$ forces the whole contribution to be zero. As a
result only the diagram that does not contain vertical impurity lines
survives in the case of the short-range impurities as it is obvious when one
considers these terms not separately but in the combined form 
$\partial /\partial p_{y}J_{y}.$
The observation about the importance of keeping combined expressions in the
calculation of $\sigma _{xy}$ is of general character. Separate terms can be
more involved than their total contribution, see, e.g., the discussion
related to $\partial ^{2}\Sigma /\partial \xi ^{2}$ following Eq.~(\ref{b_definition}).

To conclude this section we notice that  Eq.~(\ref{Fans3})  reproduces the
structure of the iterative solution of the transport equation for 
$\sigma_{xy}$. Let us compare Eq.~(\ref{Fans3})  with the term 
\mbox{$(e[\mathbf{ v}_{\mathbf{ p}}\times \mathbf{ B}]\partial /\partial \mathbf{ p})
\delta n_{\mathbf{ p}}$} that appears in the transport equation as a result of 
the action of the Lorentz force and discribes the turn of the current.
In the first bracket of Eq.~(\ref{Fans3}) the derivative 
$\partial /\partial p_{x}$ yields the velocity $\mathbf{ v}_{\mathbf{ p}}$,
while $\partial /\partial p_{y}$ in the
second bracket corresponds to the momentum derivative acting on the
distribution function $\delta n_{\mathbf{ p}}$. 
The derivation in this section was limited to the leading order 
in $1/\epsilon_{F}\tau _{tr}$.
To get further, one has to study the
dependence of $1/\tau _{tr}$ on the flux. The leading term of that kind
arises due to the Cooperon corrections and is discussed 
in Appendix~\ref{Appendix_C}.
\section{Electron-electron interaction}
\begin{figure}[h]
\centerline{\includegraphics[width=0.4\textwidth]{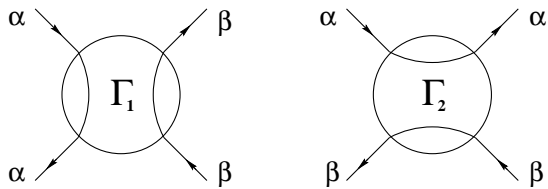}} 
\caption{ The irreducible amplitudes of the \textit{e-e} interaction $\Gamma
_{1}$ and $\Gamma _{2}$. }
\label{G1G2}
\end{figure}

\begin{figure}[h]
\centerline{\includegraphics[width=0.4\textwidth]{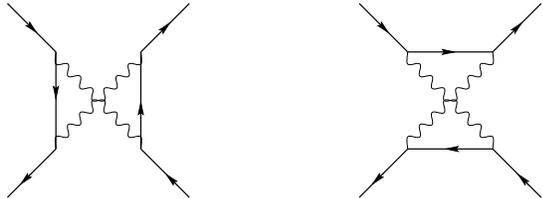}} 
\caption{ Examples of the $\Gamma _{1}$ and $\Gamma _{2}$ amplitudes.}
\label{ex_G1G2}
\end{figure}

We are interested in the current-current correlator, and therefore the
basic process to be considered is a sequential rescattering of electron-hole
pairs. Diagrammatically this process is described by the particle-hole
ladder sections alternating with the irreducible amplitudes of the 
\textit{e-e} interaction $\Gamma _{1}$ and $\Gamma _{2}$. Amplitudes $\Gamma _{1}$
and $\Gamma _{2}$ differ in spin structure as it is schematically
represented in Fig.~\ref{G1G2}. 
Examples of such amplitudes are given in Fig.~\ref{ex_G1G2}. 

A typical diagram for the Hall current correlator $\Pi _{xy}$ includes the $%
J_{x}$-current vertex, several $\Gamma $ amplitudes connected by the $R-R,$ $%
A-A$, or $R-A$ sections and, finally, the $J_{y}$-current vertex. In essence,
the correlator $\Pi _{xy}$ may be evaluated following the same line as in
the case of the long-range impurity scattering. There is, however, a
complication related to $\Gamma _{1}$ insertions. In the case of 
noninteracting electrons we have dealt with a flux of a single loop extending
from  the $J_{x}$ to the $J_{y}$ vertex. An insertion of $\Gamma _{2}$  leaves the 
number of loops in a diagram to be the same, while the $\Gamma
_{1}$ amplitude splits the loop into pieces and breaks the single loop
structure. We restore the construction with the main flux loop having in
mind to use the basic results of the previous section, i.e., 
Eqs. (\ref{Fans1}) and (\ref{Fans2}).

\begin{figure}[h]
\centerline{\includegraphics[width=0.45\textwidth]{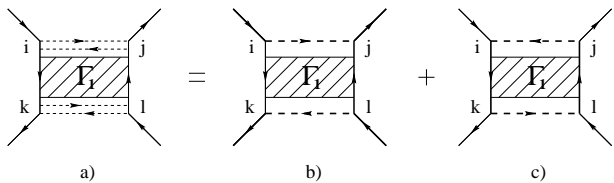}} 
\caption{ The phase factors of the $\Gamma _{1}$ amplitude. }
\label{G1}
\end{figure}

Consider an insertion of $\Gamma _{1}$ that splits a loop at the points $%
i,j,k,l.$ To restore the construction with the main flux loop we attach the
phase factor $i(e/c)(\mathbf{ r}_{j}-\mathbf{ r}_{i})\mathbf{ A}((\mathbf{ r}%
_{i}+\mathbf{ r}_{j})/2)$ to the segment $\mathbf{ r}_{i}\rightarrow \mathbf{ r}%
_{j}$ in Fig.~\ref{G1}(a). The analogous operation will be done with the
segment $\mathbf{ r}_{k}\rightarrow \mathbf{ r}_{l}.$ Simultaneously the
compensating phase factors with the opposite sign are attached to the lines
going in the opposite directions, $\mathbf{ r}_{j}\rightarrow \mathbf{ r}_{i}$
and $\mathbf{ r}_{l}\rightarrow \mathbf{ r}_{k}.$ In Fig.~\ref{G1} all these
segments are denoted by the dashed lines with arrows. These dashed lines
carry phases but do not represent any Green's functions. 
Now we regroup the phase factors. It is possible because we are interested
in $\sigma _{xy}$ that is linear in $B,$ and therefore we have to expand
and keep only linear terms in the phase factors. The phases of the two
dashed lines, $\mathbf{ r}_{i}\rightarrow \mathbf{ r}_{j}$ and $\mathbf{ r}%
_{k}\rightarrow \mathbf{ r}_{l}$, are used to obtain the flux of the main
loop extending from the $J_{x}$ to the $J_{y}$ vertex [a fragment of this
loop is presented in Fig.~\ref{G1}(b) where it is indicated by the thick
lines]. The other two phase factors attached to the segments $\mathbf{ r}%
_{j}\rightarrow \mathbf{ r}_{i}$ and $\mathbf{ r}_{l}\rightarrow \mathbf{ r}%
_{k} $ are used to form a new flux loop encircling the brick of the $\Gamma
_{1}$ amplitude in the direction opposite to the main flux loop. In Fig.~\ref
{G1}(c) all the parts of the corresponding loop are given by the thick lines.

\begin{figure}[h]
\centerline{\includegraphics[width=0.45\textwidth]{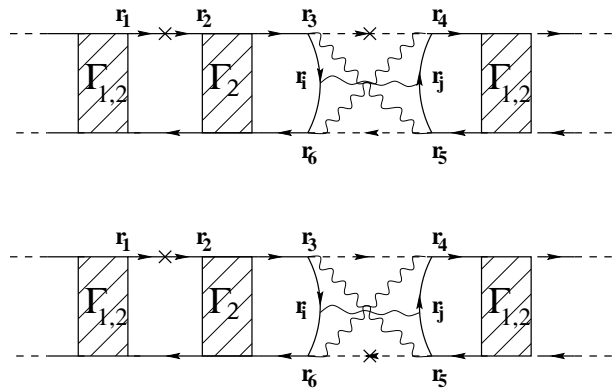}} 
\caption{ The treatment of the dashed lines in the $\Gamma _{1}$ amplitude;
the left standing derivative acts on a particle-hole section.}
\label{dashed}
\end{figure}

To get the contribution to $\Pi _{xy}$ from the main flux loop we shall
follow the derivation given in the previous section. Consider first the
situation when the left standing derivative acts on a Green's function in a
particle-hole section rather than on the $\Gamma $\ amplitudes. Let this
left standing derivative be the $p_{x}$ derivative. Then, the 
$p_{y}$ derivative is supposed to act on everything on the right of the $p_{x}$ 
derivative. The new element here is that the dashed lines added to restore
the flux structure do not carry any Green's functions, but only the phase
factors. Let us study the situation when the additional phase factors arise
from one of the $\Gamma _{1}$ amplitudes with the dashed segments ${\mathbf
r}_{3}\rightarrow \mathbf{ r}_{4}$ and $\mathbf{ r}_{5}\rightarrow \mathbf{ %
\ r}_{6}$, as shown in Fig.~\ref{dashed}. After the decomposition procedure
the corresponding flux contribution is equal to 
\begin{equation}\label{flux_dashed1}
\begin{split}
\Phi _{dashed} & =\frac{1}{2}\mathbf{ B}\cdot [ (\mathbf{ r}_{2}-\mathbf{  r}_{1})
\times (\mathbf{ r}_{4}-\mathbf{ r}_{3})
\\
 & +(\mathbf{ r}_{2}-\mathbf{ r}_{1})\times (\mathbf{ r}_{6}-\mathbf{ r}_{5}) ] .
\end{split}
\end{equation}
Next, $\Phi _{dashed}$ can be rearranged as $\frac{1}{2}\mathbf{ B}\cdot
\lbrack (\mathbf{ r}_{2}-\mathbf{ r}_{1})\times (\mathbf{ r}_{4}-\mathbf{ %
\ r}_{5})+$ $(\mathbf{ r}_{2}-\mathbf{ r}_{1})\times (\mathbf{ r}_{6}-%
\mathbf{ r}_{3})]$, and for any structure of the $\Gamma _{1}$ amplitude
one obtains 
\begin{equation}\label{flux_dashed2}
\begin{split}
\Phi _{dashed} & =\frac{1}{2}\mathbf{ B}\cdot [ (\mathbf{ r}_{2}-\mathbf{ %
\ r}_{1})\times (\mathbf{ r}_{4}-\mathbf{ r}_{j}+\mathbf{ r}_{j}-\cdots -%
\mathbf{ r}_{5})
\\
& +(\mathbf{ r}_{2}-\mathbf{ r}_{1})\times (\mathbf{ r}%
_{6}-\mathbf{ r}_{i}+\mathbf{ r}_{i}-\cdots -\mathbf{ r}_{3}) ] .
\end{split}
\end{equation}
It follows as a result of this rearrangement that after the Fourier
transformation the operation of the $p_{y}$ differentiation will move all
along the Green's functions in the segments $\mathbf{ r}_{3}\rightarrow 
\mathbf{ r}_{6}$ and $\mathbf{ r}_{5}\rightarrow \mathbf{ r}_{4}$. 

Since Eq.~(\ref{flux_dashed2}) is valid for any $\Gamma _{1}$ amplitude, we
come to the conclusion that the dashed lines do not spoil the general
structure of the discussed contributions, and every Green's function on the
``vertical'' lines, like $\mathbf{ r}_{3}\rightarrow \mathbf{ r}_{6}$ and $%
\mathbf{ r}_{5}\rightarrow \mathbf{ r}_{4}$, is $p_{y}$ differentiated on
the same footing as any other Green's function. Thus in the considered case
the $p_{y}$ derivative indeed acts on everything on the right to the $p_{x}$
derivative, as it should be expected.

Now we continue with the analysis of the case of the left standing 
$p$ derivative that acts on one of the Green's functions in the particle-hole
section. The sections $R-R$ or $A-A$ need not to be considered as each of
them yields two expressions that cancel each other out. By the $R-A$ section
we assume a set of the impurity scattering ladder diagrams that modify a
current vertex $J$ by a factor $\tau _{tr}/\tau .$ In the limit when the
external frequency $\omega \rightarrow 0$ only one $R-A$ section should be
kept in the current-current correlator $\Pi _{\alpha \beta }(\omega )$,
because each such section yields an additional power of $\omega $.
Differentiation of the Green's functions in the $R-A$ section splits the set
of the ladder diagrams into two pieces that modify both vertices $J_{x}$ and 
$J_{y}$ by $\tau _{tr}/\tau .$ As a result the contribution analogous to
that represented in Figs.~\ref{flux_dx_dy} and \ref{flux_dpy} reduces to
Eq.~(\ref{Fans1}), and in the same way the contribution analogous to Fig.~%
\ref{flux_dy_dx} reduces to Eq.~(\ref{Fans2}). Altogether we come back to
Eq.~(\ref{Fans3}) but with the current vertices and the Green's functions
dressed by the \textit{e-e} interaction.

\begin{figure}[h]
\centerline{\includegraphics[width=0.5\textwidth]{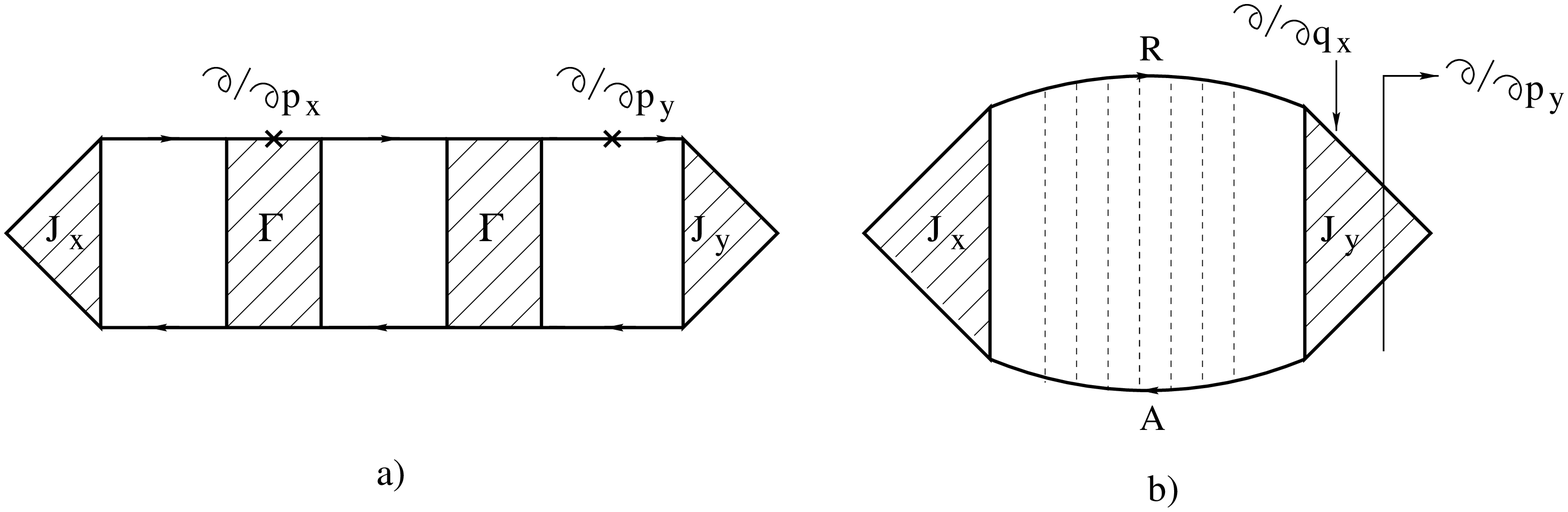}} 
\caption{ A contribution to $\protect\sigma _{xy}$ that involves 
nonpair-correlation amplitudes.}
\label{rightbar}
\end{figure}

\begin{figure}[h]
\centerline{\includegraphics[width=0.35\textwidth]{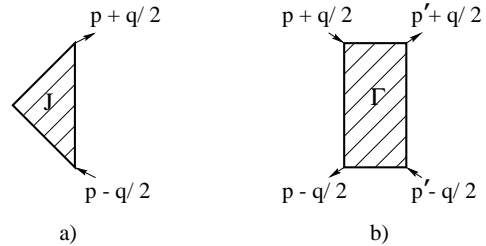}} 
\caption{ The definition of the Fourier components of the current vertexes $%
J $ and the interaction amplitude $\Gamma $. }
\label{FourierJG}
\end{figure}

The terms that have been discussed untill now determine the leading
contribution to $\sigma _{xy}\propto $ $\tau _{tr}^{2}.$ Its analysis will
be completed in the end of this section, but first we turn to the other case
when the left standing derivative, let it be the $p_{x}$ derivative, is
inside one of the interaction amplitudes $\Gamma $, while the 
$p_{y}$ derivative acts on everything to the right from this amplitude 
[an example is presented in Fig. \ref{rightbar}(a)]. As a candidate for a contribution 
of the order $\tau _{tr}^{2}$ to $\sigma _{xy}$ in the discussed process
consider the case when the $p_{y}$ derivative is applied to the $R-A$
section standing on the right from $\Gamma $. This differentiation produces
the singular terms $(\partial /\partial p_{y}{\mathcal G}^{R}){\mathcal G}^{A}$
and ${\mathcal G}^{R}(\partial /\partial p_{y}{\mathcal G}^{A})$ that
separately yield the contributions $\propto \tau _{tr}^{2}$ but in the
leading order they cancel each other out. The latter fact is obvious because
we may take the integration over $p_{y}$ by parts and then the $R-A$ section
yields a contribution that is only $\propto \tau _{tr}.$ A similar
contribution appears when the $R-A$ section is located on the left from the
amplitude $\Gamma $ that is marked by the differentiation. Arranging both
the terms in a way that the $R-A$ section becomes free from the
differentiation the contribution to $\sigma _{xy}$ in the discussed case can
be presented as 
\begin{equation}
\sigma _{xy}\propto \omega _{c}\tau _{tr}\widetilde{J}_{x}(\partial
q_{x}\Rsh \partial p_{y}-\partial q_{y}\Rsh \partial p_{x})\widetilde{J}_{y}.
\label{nonpair}
\end{equation}
Here the current vertices $\widetilde{J}_{x,y}$ are dressed by the 
\textit{e-e} interaction, but unlike $J$ they do not contain the factor $\tau
_{tr}/\tau $. The symbol $\Rsh $ [as well as the right-directed bar in 
Fig.~\ref{rightbar}(b)] 
indicates that the derivatives are ordered and the derivative over $p$ acts
on the right to the $q$ derivative where $q$ is an infinitesimal momentum in
the current vertex $J_{y}$ as shown in Fig.~\ref{FourierJG}(a). 
The momentum $q$\ has been introduced to get opposite signs for the
derivatives acting on oppositely directed lines in the current vertex. Since
the parameters of the electron liquid change on the scale of the Fermi
momentum this contribution is small compared to the leading one by the
parameter $1/\epsilon _{F}\tau _{tr}$. The reason of this smallness is that
in the discussed process the skew action of the magnetic field may 
develop only on an electron
wavelength, while for the leading term the skew effect develops on a free
path length. Notice that the processes just discussed involve the
derivatives of the $\Gamma $ amplitudes that generate terms with the 
nonpair electron correlations $f_{np}$ that were mentioned in the Introduction.
The discussed term may become significant if 
apart from the wavelength and the mean free path there is another 
length scale in the problem. 
For example, near the superconducting transition the 
fluctuations of the order parameter may contribute significantly and change 
the Hall coefficient.

The last contribution to $\Pi _{xy}$ that remains to be analyzed is related
to the flux loops encircling the $\Gamma _{1}$ amplitudes. Consider the loop
encircling the amplitude $\Gamma _{1}(\mathbf{ r}_{1},\mathbf{ r}_{2},%
\mathbf{ r}_{3},\mathbf{ r}_{4})$. The flux through the polygon ${\mathbf
r}_{1},\mathbf{ r}_{2},\mathbf{ r}_{3},\mathbf{ r}_{4}$ may be written as
the sum of the fluxes through the two triangles: 
\begin{equation} \label{fluxG2}
\begin{split}
\Phi _{\Gamma _{1}} & =\mathbf{ B}\cdot \lbrack (\mathbf{ r}_{1}-\mathbf{ r}_{2}) 
\times (\mathbf{ r}_{4}-\mathbf{ r}_{1})
\\
 & +(\mathbf{ r}_{3}-{\mathbf r}_{4})\times (\mathbf{ r}_{2}-\mathbf{ r}_{3})]. 
\end{split}
\end{equation}
We define the Fourier representation of $\Gamma _{1}(\mathbf{ r}_{1},%
\mathbf{ r}_{2},\mathbf{ r}_{3},\mathbf{ r}_{4})$ as follows 
[see Fig.~\ref{FourierJG}(b)]: 
\begin{equation} \label{Gamma2}
\begin{split}
\Gamma _{1} (\mathbf{ r}_{1} & ,\mathbf{ r}_{2}  ,\mathbf{ r}_{3},\mathbf{ r}_{4})  = 
\int \frac{d^{d}p}{(2\pi )^{d}}\int \frac{d^{d}p^{\prime }} {(2\pi)^{d}}\int \frac{d^{d}q}{(2\pi )^{d}}
\\
&  \times \exp \{i[(\mathbf{ p}+\mathbf{ q}/2) \mathbf{ r}_{1}-(\mathbf{ p}^{\prime }+\mathbf{ q}/2)\mathbf{ r}_{2}
\\
& + 
(\mathbf{ p}^{^{\prime }}-\mathbf{ q}/2)\mathbf{ r}_{3}-(\mathbf{ p}-\mathbf{ q}/2)\mathbf{ r}_{4} ] \}
\Gamma (\mathbf{ p},\mathbf{ p}^{\prime};\mathbf{ q}). 
\end{split}
\end{equation}
Comparing Eq. (\ref{fluxG2}) with Eq. (\ref{Gamma2}) we obtain that the product 
$\Gamma _{1}\Phi _{\Gamma _{1}}$ when Fourier transformed yields the
following combination that enters the correlator $\Pi _{xy}$: 
\begin{equation}
\Gamma _{1}\Phi _{\Gamma _{1}}\rightarrow \widehat{\mathbf{ B}}{\mathbf
\cdot }\left[ \frac{\partial }{\partial \mathbf{ q}}\times \left( \frac{%
\partial }{\partial \mathbf{ p}^{\prime }}-\frac{\partial }{\partial 
\mathbf{ p}}\right ) \right ]
 \Gamma _{1}(\mathbf{ p},\mathbf{ p}^{\prime };%
\mathbf{ q}).  \label{combination1}
\end{equation}
As the amplitude $\Gamma _{1}$ is taken in the limit $\mathbf{ B=0}$ it is
a scalar function of its momenta, and therefore the expression~(\ref
{combination1}) vanishes.

Summarizing the analysis presented in this section we conclude that the
leading contribution to $\sigma _{xy}$ is determined by the following
expression: 
\begin{equation}\label{leading}
\begin{split}
\Pi _{xy} & =ieB\frac{e^{2}}{2c}\left( \frac{\tau _{tr}}{\tau }\right ) ^{2}
\\
 & \times 
\left( {\mathcal G}^{A}(\mathbf{ p})\frac{\partial }{\partial p_{x}}{\mathcal %
G}^{R}(\mathbf{ p})-{\mathcal G}^{R}(\mathbf{ p})\frac{\partial }{\partial
p_{x}}{\mathcal G}^{A}(\mathbf{ p})\right )
\\
& \times
 \left( \widetilde{J}_{x}\frac{%
\partial }{\partial p_{y}}\widetilde{J}_{y}-\widetilde{J}_{y}\frac{\partial 
}{\partial p_{y}}\widetilde{J}_{x}\right).  
\end{split}
\end{equation}
The derivatives over the momentum in Eq.~(\ref{leading}) correspond to the
vector product of the two coordinate differences in the representation of
the flux in Eq.~(\ref{poly4}). The current vertexes $\widetilde{J}_{x,y}$
are not sensitive to the renormalizations by impurities due to $1/\epsilon
_{F}\tau \ll 1$, and are determined by the Ward identity:\cite{LN} 
\begin{equation}
\widetilde{J}_{\mu }=\frac{p_{\mu }}{m}\left( 1+\frac{\partial \Sigma
(p,\epsilon )}{\partial \xi _{p}}\right) ,  \label{b_definition}
\end{equation}
where $\Sigma (p,\epsilon )$ is a self-energy of the Green's function in the
presence of the \textit{e-e} interaction. 
Eqs.~(\ref{leading}) and (\ref{b_definition}) 
correspond to the solution of the transport equation for $\sigma_{xy}$ 
for the interacting electrons
[see the discussion of Eq.~(\ref{Fans3}) concluding Sec. III]. 
Due to the specific structure of
the last bracket in Eq.~(\ref{leading}) the second derivative $\partial
^{2}\Sigma /\partial \xi ^{2}$,\ which is not conventional in the 
Fermi-liquid theory, does not enter the final answer. Altogether we get the factor 
$\left( 1+\partial \Sigma (p,0)/\partial \xi _{p}\right) $ three times: two
from the vertexes and one from the $p_{x}$ derivatives of the Green's
functions in the square brackets. The integration over momentum in $\Pi
_{xy} $ can be performed keeping the Green's functions in the pole
approximation: 
\begin{equation}
{\mathcal G}^{R,A}(\epsilon ,\mathbf{ p})=\frac{(1-\partial \Sigma
(p_{F},\epsilon )/\partial \epsilon )^{-1}}{\epsilon -\frac{m}{m^{\ast }}\xi
_{p}\pm \frac{i}{2\tau }},
\end{equation}
where $\tau $ is the free path time of the quasiparticles that includes the
renormalization by the \textit{e-e} interaction, and 
\begin{equation}
m/m^{\ast }=
\left[ 1+\partial \Sigma (p,0)/\partial \xi _{p}\right]
/\left[1-\partial \Sigma (p_{F},\epsilon )/\partial \epsilon \right].
\label{m_effective}
\end{equation}
The latter relation is a standard one for the microscopical theory of the
Fermi liquid.\cite{LN} All three factors 
$ [1-\partial \Sigma(p_{F},\epsilon )/\partial \epsilon ]^{-1}$ 
created by the Green's functions
in Eq.~(\ref{leading}) together with the three factors 
$\left[ 1+\partial\Sigma (p,0)/\partial \xi _{p}\right] $
 can be combined into the cube of the
physical combination $m/m^{\ast }.$ After integration over $\xi _{p}$ one
power of $m/m^{\ast }$ will be eliminated and ultimately we obtain 
\begin{equation}
\sigma _{xy}=\omega _{c}^{\ast }\tau _{tr}\sigma _{xx},
\label{relation_star}
\end{equation}
with $\omega _{c}^{\ast }=eB/m^{\ast }c$ and 
$\sigma _{xx}=ne^2\tau_{tr}/m^{\ast }$. This result implies that no Hall coefficient
renormalization develops due to the \textit{e-e} interaction in the leading
terms over $1/\epsilon _{F}\tau _{tr}$.

\section{discussion}

We have shown microscopically that the Hall coefficient in a weak magnetic
field is not renormalized by the \textit{e-e} interaction in the
leading order in $1/\epsilon _{F}\tau _{tr}$. Although the terms that are
not conventional in the Fermi-liquid theory appear in the intermediate
stages, they do not enter the final answer. This remarkable fact
is a direct consequence of the skew structure which arises in the Hall
effect because of the magnetic flux.

The result about the cancellation of the Fermi-liquid renormalization
corrections in the Hall coefficient differs from the one obtained previously
in Refs.~\onlinecite{FEW} and \onlinecite{KOHNO}. It is formidable to make a comparison
with the analysis of these papers because the distinction of the impurity
scattering amplitudes from the \textit{e-e} interaction amplitudes was not
performed explicitly. These amplitudes have different structure in exchange
of the frequency, however. Therefore any treatment lacking clear
distinction between these amplitudes is potentially dangerous. The analysis
of the Fermi liquid in Ref.~\onlinecite{KOHNO} did not reproduce the well
established results of the transport theory, because of the confusing
treatment of the frequency integrations induced by the \textit{e-e}
interaction amplitudes. On the other hand, in Ref.~\onlinecite{FEW} the authors
were mainly concerned with the impurity scattering. One can check, in fact,
that their factor 
$[1+\partial \Sigma ^{\prime }(p,0)/\partial \epsilon(p)]^{2}$ 
in $R_{H}$ differs negligibly from unity in the case of 
noninteracting electrons scattered by a random potential.

The absence of the renormalization in the Hall coefficient corresponds to
the result that can be obtained within the phenomenological theory of the
Fermi liquid. For completeness we reproduce it here following Sec. 3.6 in
Ref.~\onlinecite{PN}. In the Fermi-liquid theory the excited states of the
interacting electrons are described by the gas of quasiparticles with the
effective Hamiltonian written via the distribution function of the
quasiparticles $n_{\mathbf{ p}}(\mathbf{ r},t)$: 
\begin{equation}
{\bar{\epsilon}}_{\mathbf{ p}}(\mathbf{ r})= \epsilon_{\mathbf{ p}}+
\sum_{\mathbf{ p^{^{\prime }}}} f_{\mathbf{ pp}^{\prime }} \delta n_{%
\mathbf{ p}^{^{\prime }}}(\mathbf{ r}).  \label{effective}
\end{equation}
Here $(\mathbf{ p}, \mathbf{ r})$ are classical variables, and the
interaction term $f_{\mathbf{ pp}^{\prime }}=\int d\mathbf{ r}^{\prime }$ $%
f(\mathbf{ pr},\mathbf{ p}^{\prime }\mathbf{ r}^{\prime })$ is assumed to
be local. In the presence of the vector potential $\mathbf{ A}(\mathbf{ r},t)$ 
one should make a transition to the pair of the conjugate variables $(\mathbf{ P}, \mathbf{ r}) $: 
\begin{equation}
\widetilde{\epsilon }_{\mathbf{ P}}(\mathbf{ r})= {\bar{\epsilon}}_{%
\mathbf{ P}-(e/c)\mathbf{ A}}(\mathbf{ r}); \hspace{20pt} \widetilde{n}_{%
\mathbf{ P}}(\mathbf{ r})= n_{\mathbf{ P}-(e/c)\mathbf{ A}}(\mathbf{ r}%
),  \label{effectiveM}
\end{equation}
where the vector potential $\mathbf{ A}(\mathbf{ r},t)$ represents both
magnetic and electric fields: $\mathbf{ B}=\nabla \times \mathbf{ A}$, $%
\mathbf{ E}=-(1/c)\partial \mathbf{ A}/\partial t$. In the phase space $(%
\mathbf{ P},\mathbf{ r})$ the flow of the quasiparticle density $%
\widetilde{n}_{\mathbf{ P}}(\mathbf{ r})$ satisfies the Liouville
equation: 
\begin{equation}
\frac{\partial \widetilde{n}}{\partial t}+ \frac{\partial \widetilde{n}}{%
\partial \mathbf{ r}} \frac{\partial \widetilde{\epsilon }}{\partial 
\mathbf{ P}}- \frac{\partial \widetilde{n}}{\partial \mathbf{ P}} \frac{%
\partial \widetilde{\epsilon }}{\partial \mathbf{ r}}=I_{coll}.
\label{Liouville}
\end{equation}
To make the dependence on the external fields explicit it is convenient to
reexpress Eq.~(\ref{Liouville}) in terms of $n(\mathbf{ p},\mathbf{ r},t)$%
: 
\begin{equation}\label{Boltzmann}
\begin{split}
\frac{\partial n}{\partial t} & + \frac{\partial {\bar{\epsilon}}}{\partial
p_{\alpha }} \frac{\partial n}{\partial r_{\alpha }}- \frac{\partial {\bar{%
\epsilon}}}{\partial r_{\alpha }} \frac{\partial n}{\partial p_{\alpha }}%
+eE_{\alpha } \frac{\partial n}{\partial p_{\alpha }}
\\
& + \frac{e}{c} \left( \frac{\partial {\bar{\epsilon}}}{\partial \mathbf{ p}}\times \mathbf{ B}
\right)_{\alpha } \frac{\partial n}{\partial p_{\alpha }}= I_{coll}.
\end{split}
\end{equation}
Next, it is useful to introduce the deviations from the global and local
equilibrium defined as 
\begin{equation}
\delta n_{\mathbf{ p}}(\mathbf{ r})= n_{\mathbf{ p}}(\mathbf{ r})-
n^{0}(\epsilon_{\mathbf{ p}}),\,\delta \overline{n}_{{\mathbf
p}}(\mathbf{ r})= n_{\mathbf{ p}}(\mathbf{ r})- n^{0}({\bar{\epsilon}}_{%
\mathbf{ p}}(\mathbf{ r})),  \label{deviations}
\end{equation}
respectively. The two are related as follows 
\begin{equation}
\delta \overline{n}_{\mathbf{ p}}(\mathbf{ r})= \delta n_{{\mathbf{ p}}}(%
\mathbf{ r})- \frac{\partial n^{0}}{\partial \epsilon _{\mathbf{ p}}}
\sum_{\mathbf{ p}^{^{\prime }}}f_{\mathbf{ pp}^{^{\prime }}} \delta n_{%
\mathbf{ p}^{^{\prime }}} (\mathbf{ r}).  \label{global_local}
\end{equation}
With the use of Eqs.~(\ref{deviations}) and (\ref{global_local}), Eq.~(\ref
{Boltzmann}) yields after linearization 
\begin{equation}
\frac{\partial \delta n}{\partial t}+eE_{\alpha }\frac{\partial n^{0}}{%
\partial p_{\alpha }}+v_{\alpha }\frac{\partial \delta \overline{n}}{%
\partial r_{\alpha }}+ \frac{e}{c}\left( \mathbf{ v}\times \mathbf{ B}%
\right)_{\alpha } \frac{\partial \delta \overline{n}}{\partial p_{\alpha }}%
=I_{coll},  \label{Boltzmann1}
\end{equation}
where $v_{\alpha }=p_{\alpha }/m^{\ast }$, and in the relaxation time
approximation $I_{coll}=-\delta \overline{n}_{\mathbf{ p}}/\tau_{tr}$. The
current density is also expressed through the deviation from the local
equilibrium 
\begin{equation}
J_{\alpha}(\mathbf{ r})= \sum_{\mathbf{ p}} \frac{p_{\alpha}}{m^{\ast}}
\delta \overline{n}_{\mathbf{ p}}.  \label{current}
\end{equation}
To study the Hall effect one should look for the stationary, homogeneous
distribution $\delta \overline{n}_{\mathbf{ p}}$. The static limit of 
Eqs.~(\ref{Boltzmann1}) and (\ref{current}) are identical to the corresponding
equations for noninteracting electrons, if $m^{\ast}\rightarrow m$ and $%
\delta {\bar n} \rightarrow \delta n$. Therefore the transport coefficients
are given by the free electron expressions with the substitution $m$ to $%
m^{\ast },$ and $\sigma _{xy}=\omega _{c}^{\ast }\tau _{tr}\sigma _{xx}$.

It is clear form this discussion that since the Fermi-liquid theory uses a
local functional with the distribution function depending on the classical
variables $(\mathbf{ p},\mathbf{ r})$ it does not contain the 
nonpair-correlation contribution~(\ref{nonpair}). The latter describes the influence
of the flux phase on the interaction amplitudes and, obviously, is beyond
the scope of the Fermi-liquid theory. Fortunately, this term is small by the
parameter $1/\epsilon_{F}\tau_{tr}$.

To summarize, we have proved that the Hall coefficient is not renormilized by the 
\textit{e-e} interaction in the leading order in $1/\epsilon_F\tau$. 
The result holds for not too low temperatures 
when the logarithmic corrections from the Altshuler-Aronov effect can be ignored, 
i.e., when $(e^{2}/ h\sigma _{xx})\ln (1/T\tau )\ll 1$. Furthemore 
it follows from the analysis of Ref.~\onlinecite{Zala} that 
while as we show the leading term is robust, 
the temperature corrections 
to the Hall coefficient remain small up to the tempertures substantially
lower than $1/\tau $.
Combined with this observation the present analysis gives an explanation of the 
stability of the Hall coefficient $R_{H}$ to the \textit{e-e} interaction 
observed in Refs.~\onlinecite{Pudalov2} and \onlinecite{Shashkin}(b) for the dilute electron 
gas in MOSFETs not too close to the metal-insulator transition. 
\begin{acknowledgments}
The authors are grateful to M. Gershenson, V. Pudalov, I. Burmistrov, 
Y. Levinson, and A. Punnoose for valuable discussions. The work was supported by
the US-Israel Binational Science Foundation (BSF)
and the Minerva Foundation. A.F. thanks the Aspen Center
for Physics for its hospitality during the completion of this work. 
\end{acknowledgments}
\appendix
\section{Symmetry properties}
\label{Appendix_A}
In this appendix the symmetry properties of the core Green's function $\tilde{G}$ are 
presented. 
A useful discussion of the symmetry properties of the Green's function
in the presence of \textit{e-e} interaction can be found in the classical 
papers\cite{Luttinger}$^{,}$\cite{Bychkov} which study the quantum oscillations in a magnetic 
field.
\subsection{Translation invariance}
A many-body system in a homogeneous magnetic field is described by the Hamiltonian
\begin{equation}
H=\sum_{i}
\frac{1}{2m}
\left(
{\bf p}_{i}-\frac{e}{c}{\bf A}({\bf r}_{i})
\right)^{2}+
\frac{1}{2}\sum_{i \neq j } V(|{\bf r}_{i}-{\bf r}_{j}|).
\label{Hamiltonian}
\end{equation}
In this case the vector potential depends linearly on the coordinates,
and we introduce a matrix $A^{\mu \nu}$ following the considerations of
Ref.~\onlinecite{Laikhtman},
\begin{equation}
A^{\mu}({\bf r})=A^{\mu \nu}r_{\nu}.
\label{A_matrix}
\end{equation}
The conjugate vector potential is defined through its transpose
$\tilde{A}^{\mu}({\bf r})=\tilde{A}^{\mu \nu}r_{\nu}$
in such a way that the gauge invariant relation
$\tilde{A}^{\mu\nu}-A^{\mu\nu}= B\varepsilon_{ \mu \nu}$ holds.
The generators of the magnetic translations are defined as
\begin{equation}
\label{generators}
T^{\mu}=\sum_{i}
\left(p^{\mu}_{i}-\frac{e}{c}\tilde{A}^{\mu}(\mathbf{  r}_{i})\right),
\end{equation}
where the sum is taken over all electrons.
Consider the operator of a finite translation
$S_{\mathbf{  a}}=\exp \left(-ia_{\mu}T^{\mu}\right)$.
The Matsubara Green's function
${\cal G}(\mathbf{  r}_{1},\mathbf{  r}_{2};\tau)=
- \langle T_{\tau}\psi(\mathbf{  r}_{1},\tau)\psi^{\dagger}(\mathbf{  r}_{2}) \rangle $
remains unchanged under the canonical transformation
$\psi \rightarrow \psi'=S_{\mathbf{  a}}^{-1}\psi S_{\mathbf{  a}} $,
as the generators (\ref{generators}) commute with the Hamiltonian~(\ref{Hamiltonian}).
Consider now this transformation explicitly. Using the Baker-Hausdorff identity
$\exp(A+B)=\exp(A)\exp(B)\exp(-[A,B]/2)$ the $S_{\mathbf{  a}} $ may be decomposed as
\begin{equation}\label{decomposition}
\begin{split}
S_{\mathbf{  a}}= & \exp\left(-i{\bf a}\cdot{\bf P}_{tot}\right)
\exp \left(i\frac{e}{c}{\bf a}\cdot{\bf \tilde{A}}_{tot}\right)
\\
& \times \exp \left[-\frac{ie}{4c} a_{\mu}a_{\nu}\left(\tilde{A}^{\nu\mu}+\tilde{A}^{\mu\nu}\right)
\right],
\end{split}
\end{equation}
where the $\mathbf{  P}_{tot}$ is the total momentum and
$\mathbf{  \tilde{A}}_{tot}$ is the sum of the conjugate vector potentials felt by all the
particles in the system.
Then the transformed operator $\psi'$ becomes
\begin{equation}
\label{transformation2}
\psi'(\mathbf{  r})=
\psi(\mathbf{  r}-\mathbf{  a})\exp
\left(\frac{ie}{c}\mathbf{  a}\mathbf{  \tilde{A}}(\mathbf{  r})\right)
\end{equation}
and correspondingly
\begin{equation}
\label{transformation3}
\psi^{'\dagger}(\mathbf{  r})=
\psi(\mathbf{  r}-\mathbf{  a})
\exp \left(-\frac{ie}{c}\mathbf{  a}\mathbf{  \tilde{A}}(\mathbf{  r})\right).
\end{equation}
Relations~(\ref{transformation2}) and (\ref{transformation3}) imply that
\begin{equation}
\label{transformation4}
{\cal G}(\mathbf{  r}_{1},\mathbf{  r}_{2};\tau)=
{\cal G}(\mathbf{  r}_{1}-\mathbf{  a},\mathbf{  r}_{2}-\mathbf{  a};\tau)
\exp \left(\frac{ie}{c}\mathbf{  a}\mathbf{  \tilde{A}}(\mathbf{  r_{1}}-\mathbf{  r_{2}})\right).
\end{equation}
It follows then from the relation~(\ref{transformation4}) that the core Green's function
defined in Eq.~(\ref{defGtilda}) is translation invariant.
In the presence of disorder the statement is still valid for averaged Green's functions.
(The average over disorder generates translation invariant two-body
``interaction'' terms
when one introduces $n \rightarrow 0$ replicas of the system, 
or uses other standard tricks.)
\subsection{Gauge invariance}
It has been shown in Ref.~\onlinecite{Luttinger} that
Green's functions are transformed as
\begin{equation}
\label{gauge1}
\exp \left(i\frac{e}{c} [\chi(\mathbf{  r}_{1})-\chi(\mathbf{  r}_{2})] \right)
{\cal G}_{\mathbf{  A}}(r_{1},r_{2};\tau)=
{\cal G}_{\mathbf{  A}+\mathbf{  \nabla}\chi}(r_{1},r_{2};\tau)
\end{equation}
under the gauge transformation
$\mathbf{  A} \rightarrow \mathbf{  A}'= \mathbf{  A}+ \mathbf{  \nabla}\chi$.
Clearly, the integral
$(ie/c)\int_{\mathbf{  r}_{2}}^{\mathbf{  r}_{1}}\mathbf{  A}d\mathbf{  r}$
in the exponential prefactor in Eq.~(\ref{defGtilda})
changes exactly in a way as to cancel out the factor appearing in the
left-hand side of Eq.~(\ref{gauge1}). 
This cancellation ensures the gauge invariance of the core
Green's function.
\subsection{Rotational invariance}
We consider the transformation properties
under rotations in the $x-y$ plane in the circular gauge,
\begin{equation}
\label{circ_gauge}
\mathbf{  A} =
\frac{1}{2}\mathbf{  B}\times{\mathbf{  r}}.
\end{equation}
In this gauge the Hamiltonian (\ref{Hamiltonian}) takes
explicitly the form which contains rotational invariant terms only.
Similarly to $S_{\mathbf{  a}}$ the operators
$S_{\mathbf{  \phi}}=\exp \left(-i\phi L_{z}\right)$
can be introduced. As $S_{\mathbf{  \phi}}$  commute with the Hamiltonian,
the Green's function remains invariant
under the action of $S_{\mathbf{  \phi}}$
which rotate the coordinates $\mathbf{  r}_{1}$ and  $\mathbf{  r}_{2}$
in ${\cal G}(\mathbf{  r}_{1},\mathbf{  r}_{2},\tau)$.
The phase factor
$(ie/c)\int_{\mathbf{  r}_{2}}^{\mathbf{  r}_{1}}\mathbf{  A}d\mathbf{  r}$
is also unchanged in the course of rotation in the circular gauge,
and therefore the core Green's function $\tilde{G}$ should be
rotational invariant as well. Since  $\tilde{G}$ is gauge invariant the statement
holds for any gauge.
\section{Spherically symmetric spectrum}
\label{Appendix_B}
Transport properties in the case of a complicated band structure are nonuniversal.
For this reason we are not going to extend the present analysis of the
renormalization of $\sigma_{xy}$ by \textit{e-e} interaction for an arbitrary spectrum.
However, the obtained result can be readily extended to $R_{H}$ in the case of
a general but spherically symmetric spectrum $\epsilon(|\mathbf{  p}|)$.
(Such spectrum can naturally appear if the degrees of freedom in some energy shell
are integrated out in the course of the Renormalization Group treatment.)
In the presence of the magnetic field the corresponding part of
the Hamiltonian is given by the gauge invariant extension  of  $\epsilon(|\mathbf{  p}|)$
with a subsequent  symmetrization required to make the Hamiltonian  Hermitian.
The latter is needed since different components of the velocity operator do not commute.
To get a symmetrized extension of the kinetic term we expand
$\epsilon(|\mathbf{  p}|)$ in the Taylor series,
\begin{equation}
\epsilon(|{\bf p}|)=\sum_{m,n}\epsilon_{m,n}p_{x}^{m}p_{y}^{n}
\label{Taylor}.
\end{equation}
The symmetrization procedure leads to
\begin{equation}\label{symmetrize}
\begin{split}
\epsilon(|{\bf p}|) & \Rightarrow
\sum_{m,n}\epsilon_{m,n} \left(C^{n}_{m+n}\right)^{-1}\sum_{P}
\left({\bf p}-\frac{e}{c}{\bf A}\right)_{P[1]}
\\
 & \times
\left({\bf p}-\frac{e}{c}{\bf A}\right)_{P[2]} \cdots
\left({\bf p}-\frac{e}{c}{\bf A}\right)_{P[m+n]},
\end{split}
\end{equation}
where the sum is taken over all inequivalent permutations $P$, 
in such a way that altogether
there are $m$ and $n$ factors $\left[{\bf p}-(e/c){\bf A}\right]_{x}$ and
$\left[{\bf p}-(e/c){\bf A}\right]_{y}$ respectively, and
$P[l]=x,y$. The current operator is determined by the variation of the energy
with respect to the vector potential:
\begin{equation}
J_{\mu}=-\delta \langle H \rangle / \delta A_{\mu}
\label{def_current}.
\end{equation}
Now we follow the main text after Eq.~(\ref{curroper1}).
Consider the diagrammatic element related to the current operator at the point ${\bf r}_{i}$
after the contractions with the operators $\psi({\bf r}_{1})$ and
$\psi^{\dagger}({\bf r}_{2})$ have been performed.
Using Eqs.~(\ref{symmetrize}) and (\ref{def_current}) and integrating by parts
we obtain for this element
\begin{equation}\label{nq_current}
\begin{split}
J^{m,n}_{\mu}  &  =\epsilon_{m,n}
\sum_{P}{\sum_{l}}^{\prime}\left(C^{n}_{m+n}\right)^{-1}
\left(-i\nabla_{{\bf r}_{1}}-\frac{e}{c}{\bf A}\right)_{P(l-1)}
\\
  \cdots  &
\left(-i\nabla_{{\bf r}_{1}}-\frac{e}{c}{\bf A}\right)_{P(1)}
G({\bf r}_{1}-{\bf r}_{i})
\left(-i\nabla_{{\bf r}_{i}}-\frac{e}{c}{\bf A}\right)_{P(l+1)}
\\
  \cdots  &
\left(-i\nabla_{{\bf r}_{i}}-\frac{e}{c}{\bf A}\right)_{P(m+n)}
G(\mathbf{  r}_{i}-\mathbf{  r}_{2}).
\end{split}
\end{equation}
Here for each permutation all terms with the projection $\mu$ are varied
to get the current $J_{\mu}$;
correspondingly the sum ${\sum}^{\prime}$ is taken over all $l$ with
$P[l]=\mu$.
In the absence of the vector potential the current operator yields
$\partial \epsilon /\partial {\bf p}$ as expected.
To obtain the diamagnetic term in its extended form [see Eq.~(\ref{diamcontr2}) above]
it is needed to apply Eq.~(\ref{defGtilda}) and to pass the
terms $[-i\mathbf{  \nabla}-(e/c)\mathbf{  A}]$ through the
phase factors using Eqs.~(\ref{commute1}) and (\ref{commute2}). \mbox{The result is}
\begin{equation}\label{nq_diam1}
\begin{split}
J_{\mu}^{m,n} & = \epsilon_{m,n} \sum_{P}{\sum_{l}}^{\prime}
( C^{n}_{m+n} )^{-1}
\exp [ ( ie/c )
(  \Phi ({\bf r}_{1},{\bf r}_{i})
\\
      +    & \Phi ({\bf r}_{i},{\bf r}_{2}) )  ]
\left(-i\nabla_{{\bf r}_{1}}- \frac{e}{2c} ({\bf r}_{1}-{\bf r}_{i} ) \times {\bf B}\right)_{P(l-1)}
\\
   \cdots &
\left(-i\nabla_{{\bf r}_{1}}-
\frac{e}{2c}\left({\bf r}_{1}-{\bf r}_{i}\right)\times {\bf B}\right)_{P(1)}
\tilde{G}({\bf r}_{1}-{\bf r}_{i})
\\
   \times &
\left(-i\nabla_{{\bf r}_{i}}+
\frac{e}{2c}\left({\bf r}_{i}-{\bf r}_{2}\right)\times {\bf B}\right)_{P(l+1)}
\\
   \cdots &
\left(-i\nabla_{{\bf r}_{i}}+
\frac{e} {2c}\left( {\bf r}_{i}-{\bf r}_{2}\right)\times {\bf B}\right)_{P(m+n)}
\tilde{G} ({\bf r}_{i}-{\bf r}_{2}).
\end{split}
\end{equation}
Here the gradients not acting on the Green's functions are canceled out due to the
symmetrization procedure.
At this stage the analysis of the main text can be extended up to the 
Eq.~(\ref{leading}),
where now
\begin{equation}
\widetilde{J}_{\mu }=\frac{\partial \epsilon(p)}{\partial p_{\mu}}
\left( 1+\frac{\partial \Sigma(p,\epsilon)}{\partial \xi _{p}}\right).
\label{nq_b_definition}
\end{equation}
The result (\ref{relation_star}) still holds with
\begin{eqnarray}
\omega_{c}^{\ast}=
\frac{\epsilon^{\prime}(p)}{p}
\left(
\frac
{1+\partial \Sigma (p,0)/\partial \xi_{p}}
{1-\partial \Sigma (p_{F} ,\epsilon)/\partial \epsilon}
\right)
\frac{eB}{c}.
\label{nq_mass}
\end{eqnarray}
It follows from Eq.~(\ref{nq_mass}) that the renormalization corrections to
$R_{H}$ are canceled out.
\section{Weak localization corrections to the Hall coefficient}
\label{Appendix_C}
The weak localization corrections originate from the quantum interference,
and diagrammatically are known to be related to a set of the ladder diagrams
in the particle-particle channel which is called ``the cooperon.'' As a
doubly charged object the cooperon in the magnetic field acquires the form
\begin{equation}
C_{B}(\mathbf{  r}_{i},\mathbf{  r}_{f})=
\exp 
\left[ 
   \left( i2e/c\right) \Phi
   \left( \mathbf{  r}_{i},\mathbf{  r}_{f}\right) 
\right]
\tilde{C}(\mathbf{  r}_{i}-\mathbf{  r}_{f}),
\label{defCtilda}
\end{equation}
where $\Phi $ is given by Eq. (\ref{phase1}) and $\tilde{C}$ has the same
symmetry properties as the corresponding function $\tilde{G}$ in
Eq.~(\ref{defGtilda}). The latter fact can be seen from the Dyson equation
for the cooperon in the magnetic field
\begin{equation}\label{Dyson_Cooperon}
\begin{split}
C_{B}(\mathbf{  r}_{i},\mathbf{  r}_{f}) & =
\frac{\delta (\mathbf{  r}_{i}-\mathbf{  r}_{f})}{2\pi \tau \nu }+
\frac{1}{2\pi \tau \nu }\int d\mathbf{  r}^{^{\prime }}
{\mathcal G}^{R}(\mathbf{  r}_{i},\mathbf{  r}^{^{\prime }})
\\
& \times 
{\mathcal G}^{A}(\mathbf{  r}_{i},\mathbf{  r}^{^{\prime }})C(\mathbf{  r}^{^{\prime }},\mathbf{ r}_{f})
\end{split}
\end{equation}
Using Eqs.~(\ref{defGtilda}) and (\ref{defCtilda}) one gets from
Eq.~(\ref{Dyson_Cooperon}) an integral equation for $\tilde{C}$,
\begin{equation}\label{Dyson_tilda}
\begin{split}
\tilde{C}  ( & \mathbf{  r}_{i},\mathbf{  r}_{f})  =
\frac{\delta (\mathbf{  r}_{i}-\mathbf{  r}_{f})}{2\pi \tau \nu }+
\frac{1}{2\pi \tau \nu }\int d\mathbf{  r}^{^{\prime }}
\tilde{{\mathcal G}}^{R}(\mathbf{  r}_{i}-\mathbf{  r}^{^{\prime }})
\\
\times &
\tilde{{\mathcal G}}^{A}(\mathbf{  r}_{i}-\mathbf{  r}^{^{\prime }})
\exp
\left[
   (2ie/c)\Phi (\mathbf{  r}_{i},\mathbf{  r}^{^{\prime }},\mathbf{  r}_{f})
\right]
\tilde{C}(\mathbf{  r}^{^{\prime }},\mathbf{  r}_{f}),
\end{split}
\end{equation}
where $\Phi (\mathbf{  r}_{1},\mathbf{  r}_{2},\mathbf{  r}_{3})$ is the
magnetic flux through the triangle with vertexes
$(\mathbf{  r}_{1},\mathbf{  r}_{2},\mathbf{  r}_{3})$:
\begin{equation}
\Phi (\mathbf{  r}_{1},\mathbf{  r}_{2},\mathbf{  r}_{3})=
\frac{1}{2}\mathbf{  B}\cdot
\left[ (\mathbf{  r}_{1}-\mathbf{  r}_{3})\times
(\mathbf{  r}_{2}-\mathbf{  r}_{1})\right] .
\label{triangle_App}
\end{equation}
Equation~(\ref{Dyson_tilda}) contains invariant ingredients only, and this
ensures that $\tilde{C}$ has the same symmetry properties as $\tilde{G}$.
One can readily derive the following equation for $\tilde{C}$:
\begin{equation}
\left[
D\left(
-i\mathbf{  \nabla}-\frac{2e\mathbf{  B}}{2c}\times\mathbf{  r}
\right)^{2}+|\omega_{n}|
\right]
\tilde{C}(\mathbf{  r},\omega_{n})=
\frac{\delta^{d}{(\mathbf{  r})}}{2\pi \tau^{2} \nu }
\label{Dyson_tildaF2}.
\end{equation}
As it has been already explained in the main text for
$\sigma _{xy}$ linear in $B$ it is enough to keep $\tilde{G}$ and
$\tilde{C}$ in the limit $B=0$.
At the vanishing magnetic field
$\tilde{C}$ turns into the singular propagator
\begin{equation}
C(\mathbf{  Q},i\omega _{n})=\frac{1}{2\pi \tau ^{2}\nu }\frac{1}{|\omega
_{n}|+DQ^{2}}.
\label{Cooperon_free}
\end{equation}

\begin{figure}[h]
\centerline{\includegraphics[width=0.45\textwidth]{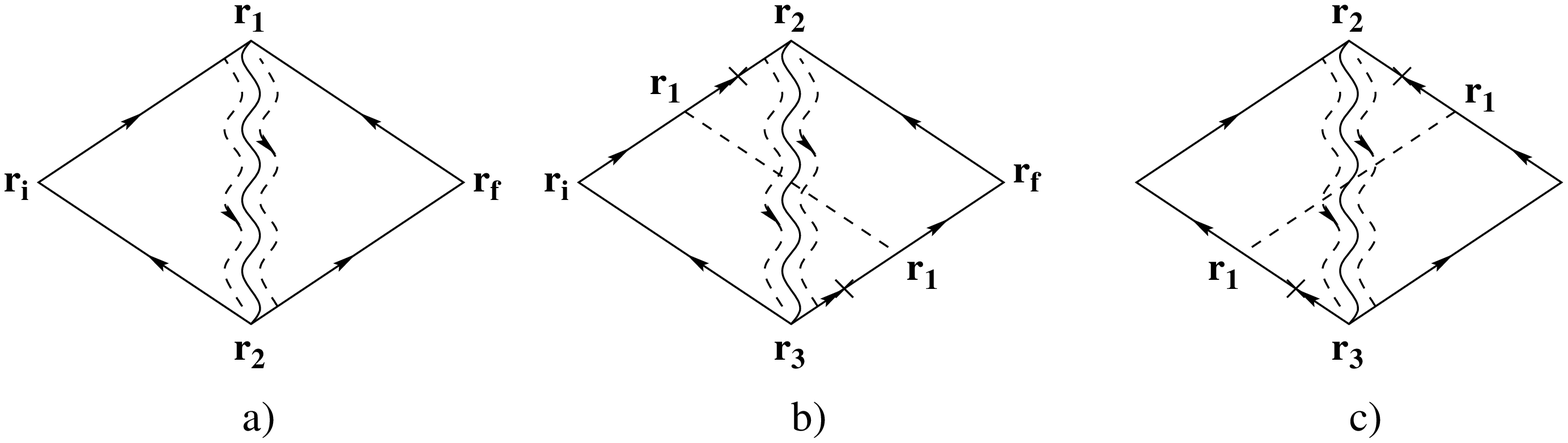}} 
\caption{Weak localization corrections to $\sigma_{xy}$.
The wavy line denotes the cooperon, the dashed wavy lines carry phases, the dashed
line denotes the scattering by impurities.}
\label{WL}
\end{figure}

In this Appendix we will assume that the scattering potential is short
ranged and that the electrons have an arbitrary but spherically symmetric
spectrum $\epsilon (p)$.
Consider first the diagram presented in Fig.~\ref{WL}(a).
With the use of Equation~(\ref{diamcontr2}) the diamagnetic contribution from the
$J_{x}$ vertex can be written as
\begin{equation}\label{xdiam_WL}
\begin{split}
       &   \Pi _{xy}^{x}    (\mathbf{  q}   =  0,i\omega _{n})=i\frac{e^{3}B}{4c}%
T\sum_{i\epsilon _{n}}\int \frac{d^{d}p}{(2\pi )^{d}}\int \frac{d^{d}Q}{(2\pi )^{d}}
\\
\times &
\varepsilon _{\mu \nu }\frac{\partial ^{2}\epsilon }{\partial p_{x}\partial p_{\mu }} 
\left( {\mathcal G}^{A}(\mathbf{  p})\frac{\partial }{\partial p_{\nu }}{\mathcal G}^{R}(\mathbf{  p})-  
{\mathcal G}^{R}(\mathbf{  p})\frac{\partial }{\partial p_{\nu }}{\mathcal G}%
^{A}(\mathbf{  p}) \right)
\\
\times &
C(\mathbf{  Q}){\mathcal G}^{R}(\mathbf{  -p+Q}){\mathcal G}^{A}
(\mathbf{  -p+Q})v_{y}\left( -p+Q\right),
\end{split}  
\end{equation}
where $v_{\alpha}(p)=\partial\epsilon/\partial p_{\alpha}$ is the velocity and
$\partial^{2}\epsilon/\partial p_{\alpha}\partial p_{\beta}$
is the inverse mass tensor. 
Together with the $J_{y}$ vertex contribution a
symmetric combination
\begin{equation*}
\begin{split}
[ ( \partial ^{2}\epsilon & /\partial p_{x}^{2}  )
( \partial \epsilon /\partial p_{y}  )^{2}
  -  2( \partial ^{2}\epsilon /\partial p_{x}\partial p_{y} )
( \partial \epsilon /\partial p_{x}  ) 
( \partial \epsilon /\partial p_{y}  ) 
\\
+ & ( \partial ^{2}\epsilon /\partial
p_{y}^{2}  ) ( \partial \epsilon /\partial p_{x} )^{2}]
\end{split}  
\end{equation*}
arises.
Exactly the same combination appears also in the calculation of
$\sigma _{xy}$ in the Drude's approximation.
In the spherically symmetric case this combination reduces to $(2/d)v_{F}^{2}m^{-1}$,
where $m^{-1}=v_{F}/p_{F}$.
[The fact that the discussed combination
contains the first derivative of $\epsilon(p)$ only was important for the derivation of
$R_{H}=1/nec$ for the arbitrary spectrum  $\epsilon(|\mathbf{  p}|)$, see
Eq.~(\ref{nq_mass}).]

To evaluate the flux term, we split the phase of the cooperon into two equal
parts indicated by the dashed lines in Fig.~\ref{WL}(a). Then the flux of this
diagram becomes a sum of the fluxes through the two triangles:
\begin{equation}
\Phi _{a}=\Phi (\mathbf{  r}_{i},\mathbf{  r}_{1},\mathbf{  r}_{2})+
\Phi (\mathbf{  r}_{f},\mathbf{  r}_{1},\mathbf{  r}_{2}).
\label{flux_a)App}
\end{equation}
When the Fourier transformation is performed the coordinate differences in
$\Phi_{a}$ lead to the differentiation with respect to momentum. To get a
nonvanishing contribution to $\sigma _{xy}$ one of the differentiation
should be applied to a current vertex. It turns out that the diamagnetic and
flux terms are identical. Together they yield
\begin{equation}
\delta \sigma _{xy}^{a}=3\omega _{c}\tau \delta \sigma _{xx},
\label{coop_a)}
\end{equation}
where $\omega _{c}=(eB/c)m^{-1}$, and $\delta \sigma _{xx}$ is the weak
localization correction to the longitudinal conductivity.

For the terms presented in Figs.~\ref{WL}(b) and (c) the diamagnetic contributions
are absent. The flux term of the diagram ~\ref{WL}(b) may be represented most
economically as
\begin{equation}
\Phi _{b}=\Phi (\mathbf{  r}_{i},\mathbf{  r}_{1},\mathbf{  r}_{3})+
\Phi (\mathbf{  r}_{f},\mathbf{  r}_{2},\mathbf{  r}_{1})-2\Phi (\mathbf{  r}_{1},
\mathbf{  r}_{3},\mathbf{  r}_{2}).
\label{flux_b)}
\end{equation}
The advantage of this representation is that the first two fluxes are
similar to the diamagnetic terms and do not contribute to $\sigma _{xy}$.
The remaining flux term leads to
\begin{widetext}
\begin{equation}\label{coop_b)}
\begin{split} 
\Pi _{xy}^{b}  (\mathbf{  q} = 0,i\omega _{n})  =  &
-i\frac{e^{3}B}{c}\left( \frac{1}{2\pi \nu \tau }\right) 
 T\sum_{i\epsilon _{n}}\int
\frac{d^{d}p}{(2\pi )^{d}}
   \int \frac{d^{d}p^{\prime }}{(2\pi )^{d}}
\int \frac{d^{d}Q}{(2\pi )^{d}}
v_{x}(p){\mathcal G}^{A}(p){\mathcal G}^{R}(p)
\frac{\partial {\mathcal G}^{A}(|\mathbf{  -p+Q|})}{\partial (-p+Q)_{x}}
\\
 &  \times 
C(Q)
{\mathcal G}^{A}
(|\mathbf{  -p^{\prime }+Q|}){\mathcal G}^{R}(|\mathbf{  -p^{\prime }+Q|})\frac{\partial
{\mathcal G}^{A}(p\mathbf{  ^{\prime }})}{\partial p_{y}^{\prime }}v_{y}(-%
\mathbf{  p}^{\prime }+\mathbf{  Q}).
\end{split} 
\end{equation}
The differentiated Green's functions are indicated by crosses in 
Figs.~\ref{WL}(b) and (c).
Similarly, the flux term of the diagram~\ref{WL}(c) is
\begin{equation}\label{coop_c)}
\begin{split} 
\Pi _{xy}^{c}(\mathbf{  q} = 0,i\omega _{n})   =   &
i\frac{e^{3}B}{c}\left( \frac{1}{2\pi \nu \tau }\right) 
T\sum_{i\epsilon _{n}}
\int \frac{d^{d}p}{(2\pi )^{d}}
\int \frac{d^{d}p^{\prime }}{(2\pi )^{d}}
\int \frac{d^{d}Q}{(2\pi )^{d}}
v_{x}(p){\mathcal G}^{A}(p){\mathcal G}^{R}(p)
\frac{\partial {\mathcal G}^{R}(|\mathbf{  -p+Q|})}{\partial (-p+Q)_{x}}
\\
    & \times
C(Q)
{\mathcal G}^{A}
(|\mathbf{  -p^{\prime }+Q|}){\mathcal G}^{R}(|\mathbf{  -p^{\prime }+Q|})\frac{\partial
{\mathcal G}^{R}(p\mathbf{  ^{\prime }})}{\partial p_{y}^{\prime }}v_{y}(-%
\mathbf{  p}^{\prime }+\mathbf{  Q}).
\end{split}
\end{equation} 
\end{widetext}
To proceed further, we rearrange the sum $\widetilde{\Pi }_{xy}=\Pi
_{xy}^{b}+\Pi _{xy}^{c}$ to the following form:
\begin{equation}
\begin{split} 
\widetilde{\Pi }_{xy}(\mathbf{ q}   &   =0,i\omega _{n})
\\
   =   &
i\frac{e^{3}B}{c}\left(\frac{1}{2\pi \nu \tau }\right) T\sum_{i\epsilon _{n}}\int \frac{dQ}{(2\pi)^{d}}
C(Q)
\\
\times &
\int \frac{d^{d}\widetilde{p}}{(2\pi )^{d}}\frac{1}{d}%
v^{2}[{\mathcal G}_{R}^{3}(\widetilde{p}){\mathcal G}_{A}(\widetilde{p})+
{\mathcal G}_{A}^{3}(\widetilde{p}){\mathcal G}_{R}(\widetilde{p}){\mathcal]}%
\\
\times &
\int \frac{d^{d}p}{(2\pi )^{d}}\frac{1}{d}v^{2}[{\mathcal G}_{R}^{3}(p)%
{\mathcal G}_{A}(p)-{\mathcal G}_{A}^{3}(p){\mathcal G}_{R}(p)].
\end{split}  
\end{equation}
One should be cautious with the last integral as 
\[
\tau ({\mathcal G}_{R}^{3}%
{\mathcal G}_{A}-{\mathcal G}_{A}^{3}{\mathcal G}_{R})=2i\xi ({\mathcal G}_{R}%
{\mathcal G}_{A})^{3}
\] 
is an odd function in $\xi .$ We have therefore to
perform an expansion in $\xi $ to get a nonvanishing result. This leads us
to an integral 
\begin{equation}\label{nonunivApp}
\begin{split}
  &  \frac{2i}{d}\int d\xi \frac{d(\nu v^{2}\tau ^{-1})}{d\xi }\xi ^{2}\left(
{\mathcal G}_{R}{\mathcal G}_{A}\right) ^{3}
\\
  &  =  \frac{4i}{d}\int d\xi \left[ (\nu
\tau ^{-1})\frac{\partial ^{2}\epsilon (p)}{\partial p^{2}}+v^{2}\frac{d(\nu
\tau ^{-1})}{d\xi }\right] \xi ^{2}\left( {\mathcal G}_{R}
{\mathcal G}_{A}\right) ^{3}.
\end{split}
\end{equation}
Unlike the expressions discussed above, the integrand in Eq. (\ref{nonunivApp})
depends on the specific form of the spectrum, i.e., it is not universal. For
$\epsilon (p)=p^{2}/2m$ and $d=2$ the diagrams of
Figs.~\ref{WL}(b) and (c) yield
$\delta \sigma _{xy}^{(b+c)}=-\omega _{c}\tau \delta \sigma _{xx}.$ Then the
weak localization correction to the Hall coefficient,
\nolinebreak[6]
\begin{equation}
\frac{\delta R_{H}}{R_{H}}=\frac{\delta \sigma _{xy}}{\sigma _{xy}}-2\frac{%
\delta \sigma _{xx}}{\sigma _{xx}},  
\label{deltaHallApp}
\end{equation}
vanishes as it was first shown by H. Fukuyama.\cite{Fukuyama2}
\nopagebreak[1000]

The possibility of an economic representation of the flux is the
essential advantage of the method.
\nopagebreak[1000]
The expressions (\ref{xdiam_WL}), (\ref{coop_b)}), and (\ref{coop_c)})
\nopagebreak[1000]
have been obtained without any intermediate steps here. 
\nopagebreak[1000]

\vspace{300pt}
\end{document}